\documentclass[journal,twoside,web]{ieeecolor}
\def\logoname{LOGO-tmi-web}

\usepackage{framed,multirow}

\usepackage{amssymb}
\usepackage{latexsym}

\usepackage{url}

\usepackage{hyperref}

\definecolor{newcolor}{rgb}{.8,.349,.1}
\definecolor{subsectioncolor}{rgb}{0.03,0.45,0.6}

\usepackage[linesnumbered,noline,noend]{algorithm2e}
\usepackage{graphicx}
\usepackage{subfig}
\usepackage{amsmath}
\usepackage{adjustbox}
\usepackage{float}
\usepackage[table]{xcolor}
\usepackage{multirow}
\usepackage{sidecap}
\usepackage{wrapfig}
\usepackage{array}
\usepackage{color}
\usepackage{colortbl}
\usepackage{hhline}
\usepackage{verbatim}
\usepackage[switch]{lineno}

\newcommand{\PV}{PV}

\newcommand{\pathtrained}{PTN}
\newcommand{\spheretrained}{STN}
\newcommand{\cartoThree}{CARTO\textsuperscript \textregistered \space 3 System\space}
\newcommand{\sfcath}{THERMOCOOL SMARTTOUCH \textsuperscript \textregistered SF Catheter}
\newcommand{\lassocath}{LASSO \textsuperscript \textregistered Catheter}

\newcommand\addtag{\refstepcounter{equation}\tag{\theequation}}
\newcommand{\vnet}{V-Net}
\newcommand{\ded}{DED}
\newcommand{\vnetsp}{V-Net }
\newcommand{\dedsp}{DED }

\newcommand{\journalname}{Preprint}

\def\BibTeX{{\rm B\kern-.05em{\sc i\kern-.025em b}\kern-.08em
    T\kern-.1667em\lower.7ex\hbox{E}\kern-.125emX}}

\begin{document}

\bstctlcite{BSTcontrol}
\author{Alon Baram, Moshe Safran, Tomer Noy, Naveh Geri and Hayit Greenspan,\IEEEmembership{Member, IEEE}
\thanks{ Baram. A (e-mail: alontbst@gmail.com) and Greenspan, H, are with Tel Aviv University, Faculty of Engineering, Department of Biomedical Engineering, Medical Image Processing Laboratory, Tel Aviv 69978, Israel.}
\thanks{Greenspan, H, is currently affiliated with the Department of Radiology, Icahn School of Medicine,  Mount Sinai, New York, NY, USA}
\thanks{Tomer Noy and Naveh Geri, are with Tel Aviv University, Faculty of Engineering, Department of Biomedical Engineering, Medical Image Processing Laboratory, Tel Aviv 69978, Israel.}
\thanks{Safran M., is with RSIP Vision, 16 King George, Jerusalem 94229, Israel}
\thanks{This study was partially supported by a grant from the Nicholas and Elizabeth Slezak Super Center for Cardiac Research and Biomedical Engineering at Tel Aviv University.}}

\title{Neural Network Reconstruction of the Left Atrium using  Sparse Catheter Paths }


%
%
%
%
%

\maketitle
\begin{abstract}
Catheter based radiofrequency  ablation for pulmonary vein isolation has become the first line of treatment for atrial fibrillation in recent years.
This requires a rather accurate map of the left atrial sub-endocardial surface including the ostia of the pulmonary veins, which requires dense sampling of the surface and takes more than 10 minutes. The focus of this work is to provide left atrial visualization early in the procedure to ease procedure complexity and enable further workflows, such as using catheters that have difficulty sampling the surface. 
We propose a dense encoder-decoder network with a novel regularization term to reconstruct the shape of the left atrium from partial data
which is derived from simple catheter maneuvers. To train the network, we acquire a large dataset of 3D atria shapes and generate corresponding catheter trajectories. 
 Once trained, we show that the suggested network can sufficiently approximate the atrium shape based on a given trajectory. 
We compare several network solutions for the 3D atrium reconstruction.  We demonstrate that the solution proposed produces realistic visualization using partial acquisition within a 3-minute time interval. Synthetic and human clinical cases are shown.
\end{abstract}
%
\section{Introduction}
\label{S:1}
\subsection{Clinical Problem}

\begin{figure}
 \centering
 \subfloat[]{\includegraphics[scale = 0.22]{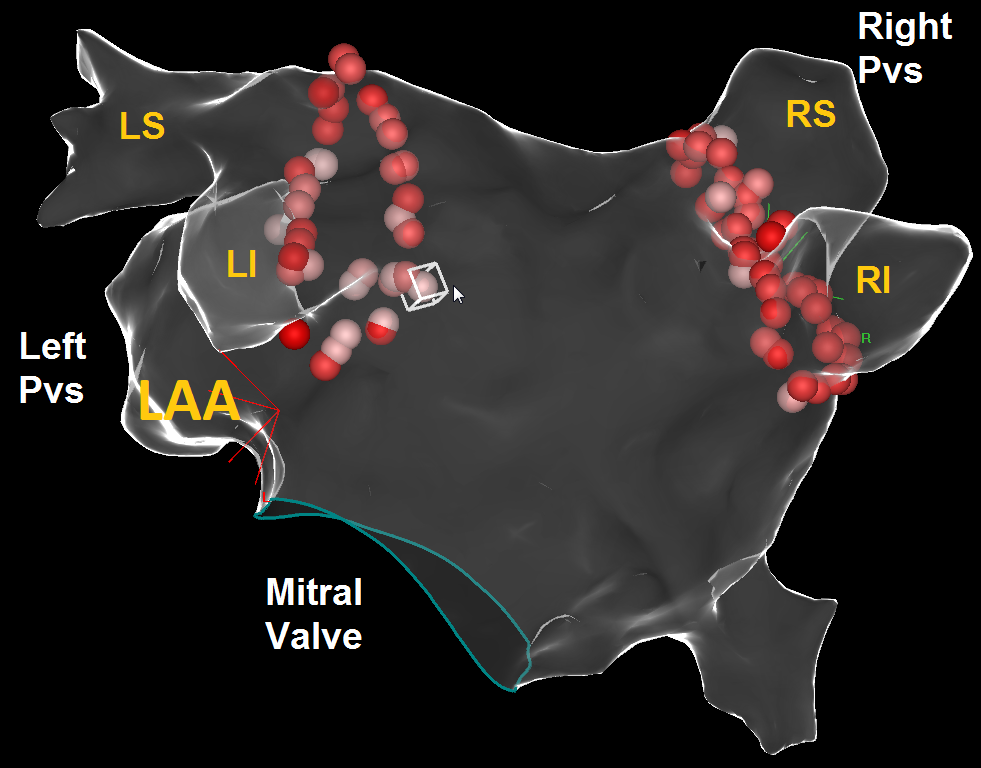}\label{laPhysb}} 
     	\quad  
  \subfloat[]{\includegraphics[scale = 0.3]{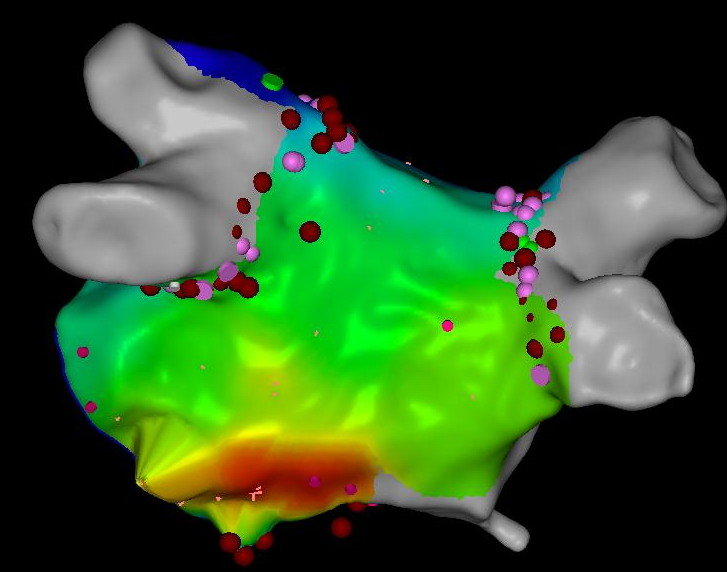}\label{laPhysa}} 
    \qquad
     	
  \subfloat[]{\includegraphics[scale = 0.16]{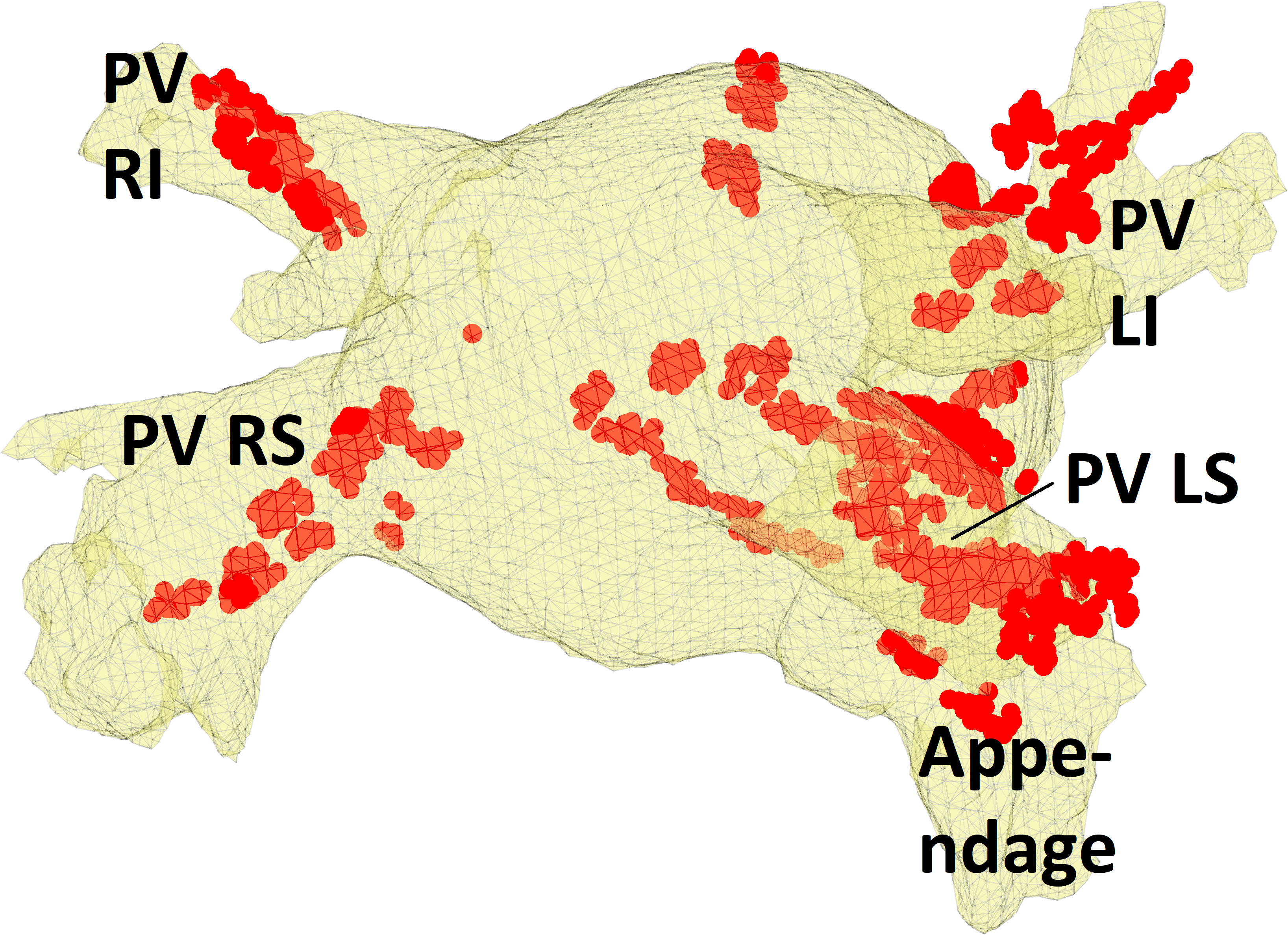}\label{fig:ct}} 
 	\caption{Clinical use of LA surface. (a) LA anatomy and typical ablation (red points), (b) LA electric propagation map. (c) Initial bearing path in red, CT anatomy is in yellow. Most of the red points lie inside the blood-pool (not over the surface).  }
    \label{laPhys} 
 \end{figure}

Atrial fibrillation (AF), the most common form of cardiac arrhythmia in humans, affects millions of people worldwide annually. It is associated with increased risk of embolic stroke
and decreased quality of life   \cite{Calkins20182017FibrillationShort}.   Catheter based electro-anatomic-mapping (EAM) 3D guided radiofrequency ablation for pulmonary vein isolation (PVI) is rapidly becoming the first line of treatment. There are many other types of arrhythmia that are treated using catheter ablation some of which may appear during or after a PVI treatment for a significant portion of the cases,  \cite{veenhuyzen2009atrial}.
EAM systems record position and electrical signals acquired as the catheter is moved inside  
the chamber. Once a large enough number of points are accumulated, a geometric reconstruction algorithm reconstructs the endocardial surface. Using local activation times, the electrical propagation wave (over the surface) that causes chamber contraction is approximated. 
The procedure is schematically depicted in Figure \ref{laPhys}.  The system provides visualization of the left atrial (LA) endocardial surface anatomy including the pulmonary veins (\PV{}) anatomical parts (LS - left superior,  RS - right superior, LI - left inferior, RI - right inferior, LAA - left atrial appendage).
Current EAM systems, accumulate the locations traversed during the catheter's trip inside the LA and produce the surrounding shape as the boundary surface. The boundary extraction process is referred to as fast anatomical mapping (FAM) \cite{sciarra2010utilityshort}. Producing an accurate anatomically correct LA surface requires the catheter to touch a large portion of the boundary (extracting many surface points). This requires extensive maneuvering of the catheter which requires physician skill, since little visual guidance is available, and takes tens of minutes of procedure time.  The ablation itself is supported by local electrograms, force indications, and various indices showing whether the catheter is ablating in the right location \cite{SKALA2015e470}.  Some anatomical regions require special care or should be avoided completely when ablating such as the esophagus or deep inside the PVs \cite{CAPPATO20091798}.

Various imaging modalities such as magnetic resonance imaging (MRI), intra-cardiac ultrasonic catheters, etc. can be used for capturing the anatomical shape by segmenting the surface of the boundary from the acquired image data.
Imaging systems must address issues such as acquisition time, radiation exposure, limited field of view (e.g. partially visited areas), noise and contrast, as well as the deformation of the heart shape due to breathing, heartbeat, and pose, in order to estimate the shape.
A typical approach to segmentation is to transform the acquired image into a 3D map indicating the probability that a point belongs to the tissue or the blood pool.
The resultant segmentation must comply with the constraint that the extracted shape  be smooth. Prior knowledge of various  anatomical details, must also be integrated.

 Anatomical imaging guidance can vary in scope and performance. Though findings are inconclusive as to the benefits of imaging methods (CT/MRI) in terms of the clinical outcomes of EP procedures, they do help reduce cognitive load, simplify the procedure and enable less proficient surgeons to achieve better results, \cite{liu2012impact}. The accurate imaging of difficult to reach areas of the atrium such as the ridge between the left superior vein and the left atrial appendage improve catheter navigation and the visualization of tissue contact during the actual ablation procedure \cite{Borlich2018-ok}.

Physicians can use different types of workflows when treating patients, as a function of the arrhythmia, the systems, and catheters available, level of proficiency, site regulations, etc. There are various types of catheters, each of which has a different shape and different parameters depending on whether it is used for diagnostic purposes or ablation, the type of arrhythmia, the patient, anatomy, etc. This affects catheter maneuverability, its ability to touch while not bending the surface, and the parts of the catheter whose location is tracked. It also directly impacts the accuracy of anatomical maps. Recently, single-shot catheters (large spherical 'balloon' like) that can ablate a PV within seconds have become popular. However, they require guidance via fluoroscopy, CT/MRI, or the insertion of additional mapping catheters \cite{Schilling2021-lf} \cite{Aryana2016}. A common workflow for a physician's initial bearing is traversing a path, namely 'initial bearing path', that traverses known anatomical landmarks such as the four pulmonary veins ostia, and is acquired in under three minutes. An example of this path, alongside the corresponding anatomical shape can be seen in Figure \ref{fig:ct}.


%

In the current study, we address the task of mapping the endo-cardial surface of the LA using a {\it{portion}} of the catheter traversal path, to provide the physician with {\it{early visualization}}, and reduce the mapping time while maintaining anatomical accuracy, especially in the PV ostia. This can also provide guidance for catheters that have difficulties sampling the surface such as balloon catheters. Our system is trained for the LA anatomy variation with $4$ \PV s \space which corresponds to the majority of the population. 
The openings and orientations of the \PV s \space reconstructed by our system should have minimal errors, and the anatomical parts should be easily identifiable. 

\subsection{Proposed Solution}
Our method is composed of two main steps, a path generation step, and an atria-shape reconstruction step. These are depicted in Figure \ref{fig:blockdiag} and described briefly next. In the first step, we generate paths resembling the path of the initial bearing to create a dataset for training. Note that the sequence of the acquired path is not considered, only the spatial locations of the points. Those are the input to a neural network whose goal is to provide anatomy surface reconstruction of the LA according to the criteria mentioned above. Figure \ref{fig:blockdiag} and Section \ref{sec:sysover} describe the system, following is a brief overview.

Supervised neural network-based algorithms require large tagged data sets. in our case, the training set should consist of meaningful relevant samples that allow for de-noising and the reconstruction of the LA shape from partial,  noisy information. Unfortunately, a large enough set of patient atrial data, usually from CT or MRI, is not easily accessible. 
In order to address this issue, we use a Left-Atrium  generator, developed by Biosense Webster, that can generate instances of LA anatomical shapes \cite{safran2017model}. 
 In the current work, we develop a novel algorithm to create clinically feasible simulated sparse catheter paths, for each such generated LA shape. 
 We show in the experiments that the path generation process provides a more robust solution for real-world catheter path scenarios.

In a follow-up reconstruction task, we aim to reconstruct the LA shape from a given catheter path input. We use a network-based solution, that receives a corrupted data-set of points as its input and is trained to predict the original, uncorrupted full data-set of points as its output. The network we use is a dense encoder-decoder network (hereon referred to as the \dedsp solution), as shown in Figure \ref{fig:blockdiag}. We compare it to the prominent \vnet \space network \cite{Milletari2016V-Net:Segmentation},  and a basic mean shape solution. The \ded \space loss incorporates a novel regularization term that enables the learning of smooth shapes. 
Since the catheter paths performed in the clinic differ from the synthetic ones, an enhanced surface boundary loss is introduced, improving solution robustness.

In the experiments, we include both model-based simulations as well as clinical human cases. We show that the proposed \dedsp  network accuracy is $5mm $. This outperforms a standard mean-shape solution in which the mean atria is rigidly transformed to match the input. We also demonstrate that the proposed solution provides anatomically relevant outputs, whereas the \vnet \space architecture solution does not. Finally, The mapping time required for the reconstruction is less than $3$ minutes. This enables a shape visualization at a much earlier time in the procedure as compared to existing solutions.



The main contributions of this study include:
\begin{itemize}
    \item Generation of a training dataset that is comprised of   LA shapes and corresponding paths.  
     \item Reconstruction methodology that takes a partial catheter path as input and is able to reconstruct a clinically viable LA shape as its output. Novel input augmentation and network loss functions are used 
     to increase the robustness of the solution to the noisy (partial) synthetic catheter paths used in the training process. 
    \item The overall solution proposed provides early visualization and guidance even for catheters that are difficult to map with.
\end{itemize}
This study is an extension of a preliminary conference version \cite{abaramMiccai2018}.
The structure of the paper is as follows:  Section \ref{sec:related_work} overviews relevant literature for this work.
Section \ref{methods:main} describes the problem definition and system structure, data generation method, and network losses. Our experiments are detailed in Section \ref{sec:expriments}, which details the chosen network architectures, and evaluation methods. Results for the experiments over the test set and the clinical cases are then provided.  
We discuss the current results in Section \ref{sec:discussions} and conclude in Section \ref{sec:conclusions}.
\section{Related Work}
\label{sec:related_work}
\subsection{Anatomical Mapping}
A given EAM system uses the locations traversed by the catheter in order to create the anatomy surface.
The authors in \cite{sciarra2010utilityshort} examined the efficiency of the \cartoThree \space fast anatomical mapping (FAM) in $25$ patients undergoing PVI procedures. They used a \lassocath \space circular multi-electrode catheter to map the LA. Each FAM point was defined as the average of the catheter locations in a one second respiratory gated window. Once enough points were acquired the surface was reconstructed using an alpha shape algorithm \cite{alphashapes}. Measurement indicated that mapping with FAM took $9\pm 3$ minutes. The accuracy of the surface was compared to an MRI scan and yielded a mean distance of $3.46\pm 0.02$mm. This gave a vein isolation success rate of $96$ percent. Note that for many cases the quality of FAM after $10$ minutes is not satisfactory; thus often physicians take more than 20 minutes to make the mapping, and furthermore,  manual shape-editing steps are required following the mapping. 

In recent work by Biosense Webster \cite{safran2017model},   termed ``model FAM (mFam)",  
a blending of shape models is used with learned statistical modeling of shape parameter distributions. This model initiates with a sparse catheter path (similar to our considered input)  and provides LA reconstruction. Later in the procedure, the reconstruction is improved using additional physician inputs and accumulated catheter data. The method shows promising results, however, so far it has been  evaluated only by visual inspection of the physicians \cite{schwartz2021reconstruction}.
\subsection{Neural Networks for 3D Shape Representation}
 Works can be found in the literature that shows success in reconstructing 3D shapes from partial data using  generative networks, \cite{3Drepreview2018}.  In \cite{brock2016generative}, the authors generated 3D volumes of different shapes and interpolated them using a variational auto-encoder combined with a fully convolutional neural network. 
 Several works focus on segmenting and reconstructing heart chambers, with either 2D images as input or 3D volumes, \cite{Cardiac3DSegReview}.
  In \cite{avendi2016combined},  an auto-encoder was used to train a 2D CNN to detect and segment the left ventricle in an MRI image.  
Recently, \cite{Liao2018MoreContouring} used several ultrasound views, acquired using an intracardiac catheter, as a 3D input to a \vnet \cite{Milletari2016V-Net:Segmentation} that inferred the complete LA shape. The \vnet \space was trained using GAN. A second network used the complete 3D shape information to improve the LA segmentation of the ultrasound slices.  
In \cite{Xiong2022}  a \vnet \space variant is used to reconstruct the LA shape using point clouds acquired by clinical mapping systems. 
This work is one of the first that combines the network solutions within this task domain. In the approach suggested, 20\% to 40\% of the atria surface is sampled, to produce the point cloud sample input to the network. Strong dice scores and surface-to-surface distance are presented. 
In the current work, we are approaching the task of the LA shape reconstruction from a different input perspective: we focus on a rapidly-acquired catheter path that traverses anatomical landmarks only.
Most of this path is in the blood-pool and we do not require it to touch any portion of the surface.

\section{Methods}
\label{methods:main}
\begin{figure*}
 \centering

  \includegraphics[scale = 0.235]{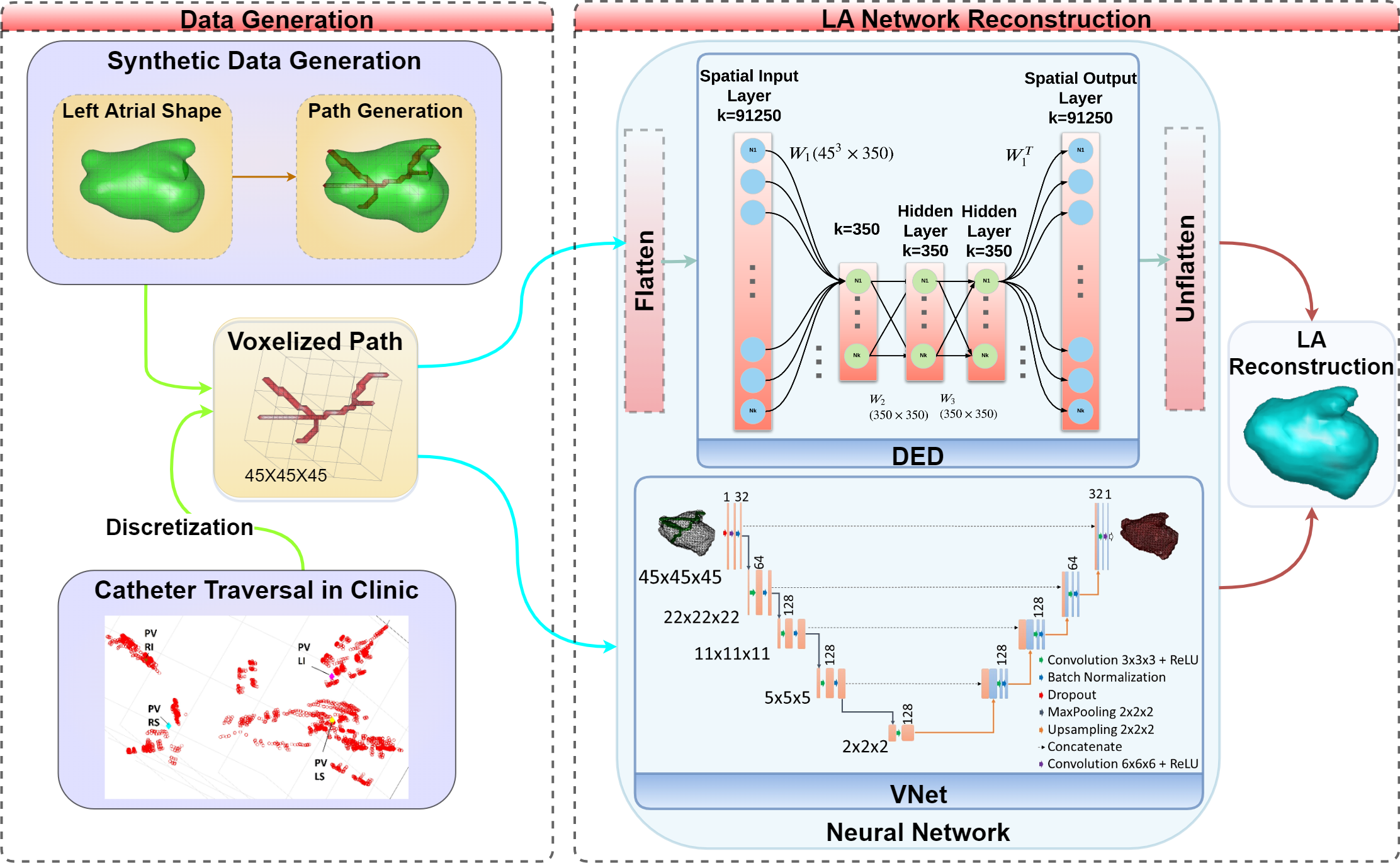}   	
 	\caption{System Block Diagram: in the data generation step  (left), a path is obtained via either  synthetic generation or by a clinical acquisition. For generating the training data, the LA shape is obtained from CT or model then the Synthetic path is generated (in red), and passed to the network; In a clinical setting, a point cloud is sampled by the catheter, registered, and discretized; The network (\dedsp or \vnet) outputs the most probable atrium (right). The parameters for both network architectures are shown.} 
    \label{fig:blockdiag}
 \end{figure*}

This Section describes our framework which generates realistic sparse data from LA shapes and completes the full shape using neural networks. 
\label{sec:sysover}
The proposed system teaches a learning algorithm to reconstruct the shape of the left atrium from a catheter path. Due to the limited number of ground-truth CT shapes with corresponding catheter paths, we simulated this data using the Biosense LA instance generator \cite{safran2017model} and a catheter path generator to train the model.  Figure \ref{fig:blockdiag} shows the data generation phase (left), where we create a sample of LA and a synthetic catheter path inside it. 
The  catheter trajectory is the input to the network  which reconstructs the LA surface (right). 
In the training process, a path, represented by a binary occupancy volume, which would closely resemble the actual path in a typical clinical scenario, is generated using given anatomical landmarks.
In the clinical case, a point cloud is acquired by a catheter and then discretized into a volume.
The result is a volume where a voxel indicates whether the catheter traversed it as depicted in Figure \ref{fig:blockdiag}. The atrium instance sampling process is discussed in Section \ref{sec:recscen}, The path is created using a graph-based algorithm, as described in Section \ref{sec:pathgen}.
In the reconstruction phase, this path is uploaded to a neural network such as the encoder-decoder network and the \vnet \space to reconstruct the original LA shape, as explained in Section \ref{sec:larec} and Section \ref{lbl:vnet}, respectively.

 \subsection{Input Data Generation: Synthetic Atria}
\label{sec:recscen}
 

 We represent the shape data as a binary 3D volume of size $45^3$ voxels, where each voxel represents a volume of $2.666mm^3$. The volume describes a set where each voxel is assigned the value of one if it is inside or on the boundary of the set, and zero otherwise, namely an occupancy volume. 
 We use a predefined model for an atrium shape \cite{safran2017model}: An atrium is defined as a blending of parametric tube-like shapes that undergo a non-linear transformation to create the atrial shape. A statistical model based on features such as \PV \space positions, \PV \space orientations, and ridge locations is modeled as a multivariate normal distribution (MVN) whose parameters are learned from CT scans. A generated sample from the model is given a statistical score using the model of how likely it is to represent an atrium.
 
 
 To generate a synthetic left atria we sampled from the MVN of the model parameters. Then, we keep samples that score well within the statistical model. Our generated dataset contains $5006$ and $1800$ samples for training and testing, respectively.

\subsection{Input Data Generation: Generating Synthetic Paths}
\label{sec:pathgen}
\begin{figure}
 	\centering
 	\subfloat[]{\includegraphics[scale=0.07]{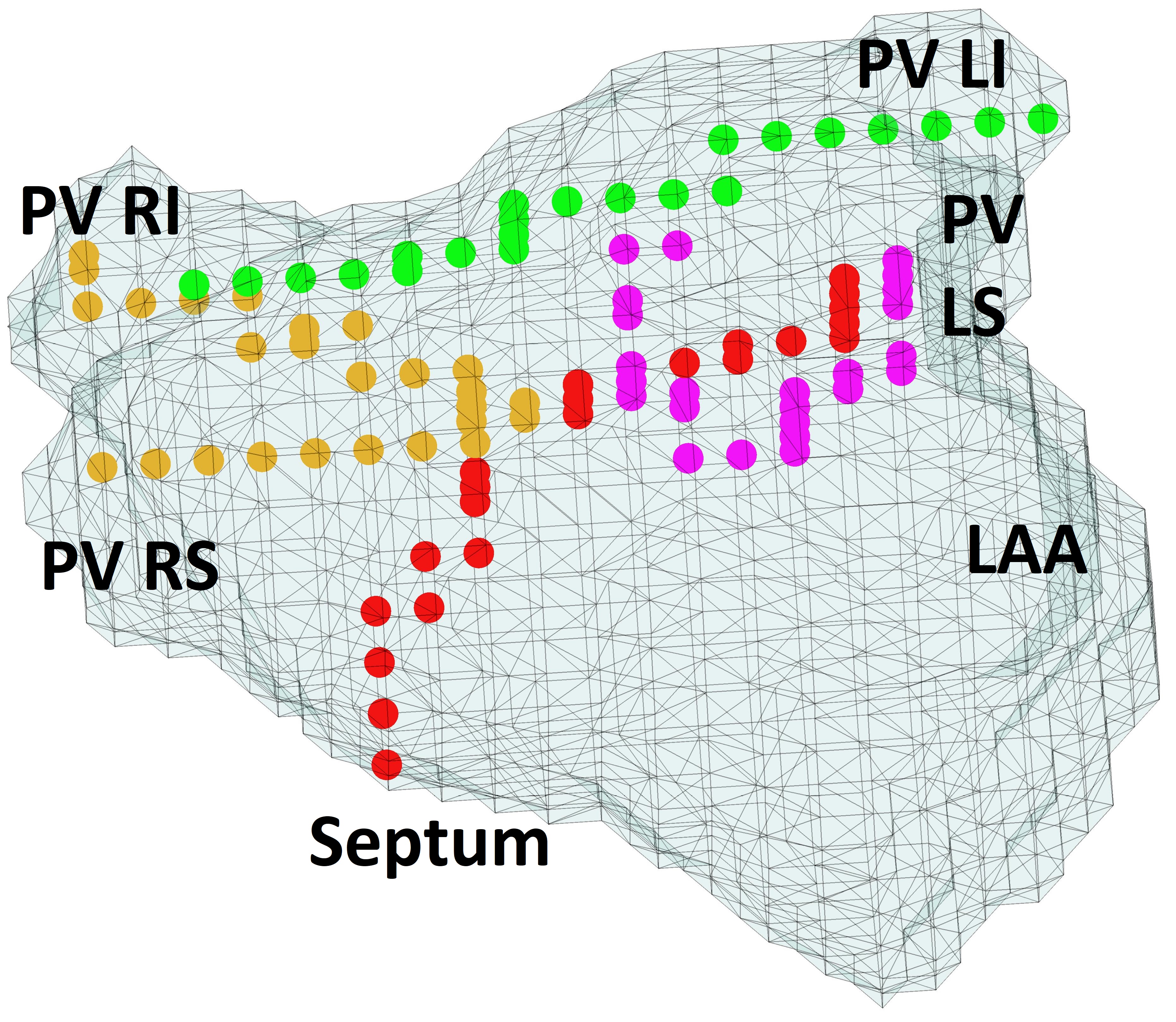}\label{fig:pathpartscol}} \quad
 	\subfloat[]  
{\includegraphics[scale=0.35]{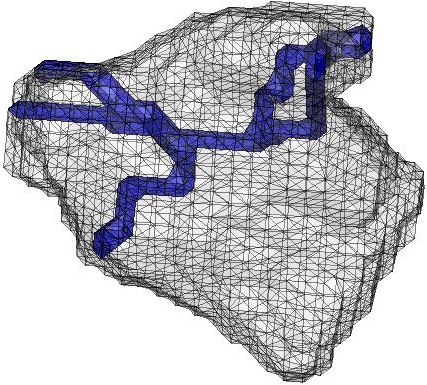} \label{fig:examplepath}}
\quad
  \subfloat[]{\includegraphics[scale = 0.43]{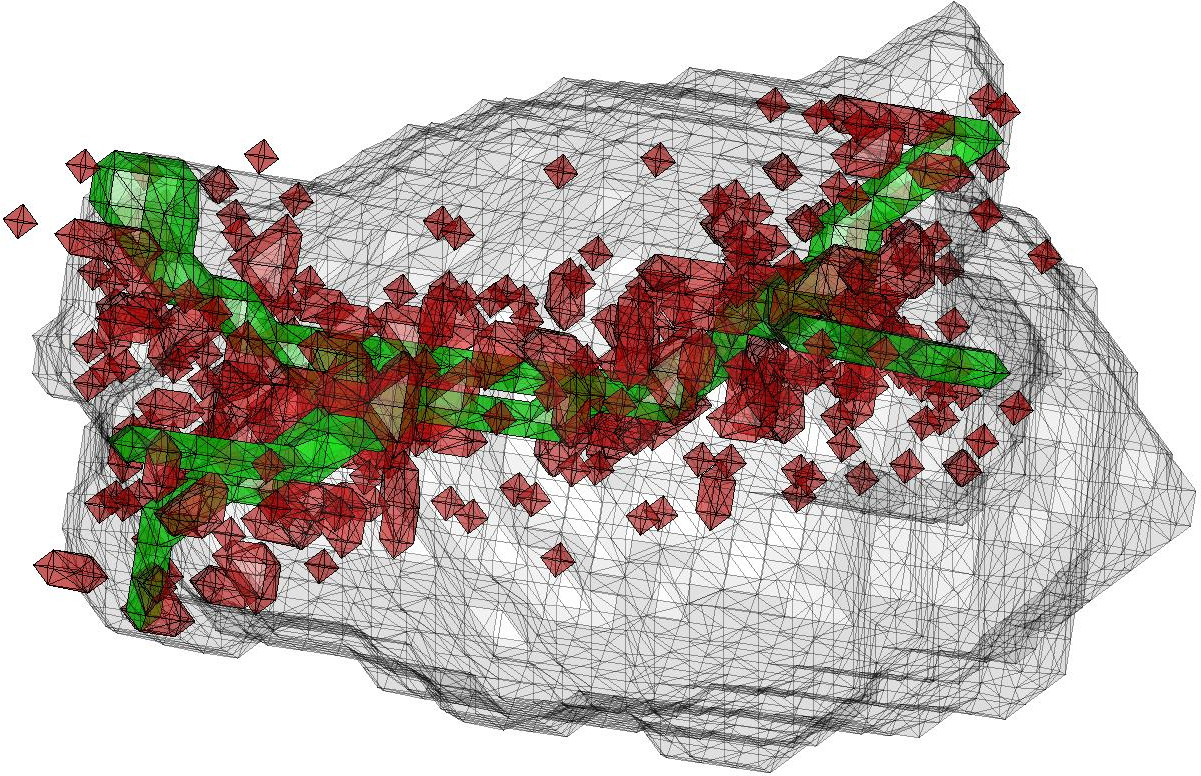}\label{fig:pathaug} }  	
 	\caption{Synthetic path creation process. (a) Each color represents a traversal from one ostia to another. Septum to PVLS - red, PVLS to PVLI - purple, PVLI to PVRI - green, and PVRI to PVRS - brown. (b) A path composed of a concatenation of the traversals in (a). (c) The green-composed path was augmented using random sampling around it to produce the red points.}
    \label{fig:simulated path}
 \end{figure}

 Next we describe the algorithm to generate a path inside  an input LA shape. The path must closely resemble the actual catheter traversal performed in the clinic. We assume a catheter with a single mapping sensor for this application. The path is generated in the following sequence: starting at the septum (the entry point from the right to the left atrium), the path proceeds to the left superior, left inferior, right inferior and finally to the right superior \PV s. Each such step represents a certain section in the overall path.
 Figure \ref{fig:simulated path} shows different path sections (a) and the overall composed path (b).  The proposed algorithm is capable of generating a multitude of simulated paths in the large group of simulated left atria in our generated dataset. 
 
The procedure to create a path is as follows: First, we locate a point in the ostium (entry point) of each \PV. Then, applying a predefined traversal order, we find a path from each ostium to the other by solving a graph optimization problem. The graph formulation trades off the shortest path distance with navigational feasibility in the clinic. We solve the graph problem by using the Dijkstra algorithm \cite{cormen}. We next review the main steps in the path creation. Additional detail  is provided in the Appendix.

\subsubsection{Simulated Paths Creation Procedure Overview}
The input LA shape for this algorithm is represented as a triangulated mesh. The mesh is a collection of vertices and faces (triangular) that describe the surface boundary of the atrium. In a preliminary stage, an Atria {\it{mean shape}} was generated by taking a voxel-wise average over all the generated LA used for training.
We manually chose the landmark coordinates of the mean shape which are the four ostia points (at the center of the ostium) and approximate septal point. 
For each new input, we located the \PV s' ostia points and the Septum. For this, the mean shape landmark points were projected to the current sample via the algorithm described in Appendix  \ref{sec:synpathstg1}.
The input is then converted into a voxel-based representation by sampling the points of the coordinate grid that are inside the mesh surface (assigned a value of one) or outside ( zero valued). The sampling resolution for the path generation is $2.666mm^3$ per voxel in a grid of $45^3$ voxels. In total, this covers a volume of $12cm^3$, which contains most atria. 
We define a trajectory as a sequence of adjacent voxels originating in the septum through the four \PV s' ostia points. Due to the small number of degrees of freedom of the catheter, the trajectory seldom follows a straight line. A realistic path tends toward the center of the atrium before reaching the next target point. Appendix \ref{sec:ptppath} describes this formulation.  
As an outcome of the procedure above, the algorithm returns a volume where marked voxels (with value one) lie on the path, and zero otherwise.

\subsubsection{Synthetic Path Augmentation}
The generated synthetic path was augmented by adding nearby points which are mostly inside the corresponding atrium. The augmentation procedure is motivated by the catheter's physical setting and was tested empirically. The catheter vibrates during movement and may exceed the chamber boundary by slightly pushing it. First, for each (grid sampled) path point $x_p$ we sample $n$ points normally distributed around it, $x_n\sim N(x_p,\sigma)$. Then the points are trimmed using a probability factor $s_f$, for each point. Next, we only consider points that are interior to the ground truth mesh. These undergo a normally distributed translation $t_f \sim N(0,1)\cdot \mu_s$, where $\mu_s$ is the factor that determines the noise level. Figure \ref{fig:pathaug} shows an augmentation result.   

\subsection{Left Atrial Shape Reconstruction using \ded \space Network}
\label{sec:larec}
We next focus on the reconstruction part of this work. We propose an NN-based reconstruction of the complete LA shape from a given sparse catheter path.
A given network is trained on synthetic paths. 
The output of the network is a probability volume, resulting from a sigmoid function applied in the last layer. This layer represents the probability of each voxel being the interior (or on the boundary) of the atrium. The value of each voxel is converted to a binary value by setting a $0.5$ threshold. 
 
 In this study, we chose the dense-encoder-decoder (\ded) as our model of choice. 
 This architecture is based on the auto-encoder (AE), which is a neural network that performs  non-linear dimensionality reduction,   similar to the non-linear (kernel) principal component analysis (PCA).  A binary vector input $x \in \left\{ 0,1 \right\}^{s}$ is mapped  to a hidden representation $y \in \left[ 0,1 \right]^d$ ,  $d \ll s$. A deterministic mapping $y=\sigma \left(Wx+b\right)$, is used, where $W:s\rightarrow d$ is an $[ d\times s ]$  matrix and $b\in R^d$  and $\sigma$ is a non-linear squashing function such as a sigmoid or a tanh.
 A decoding layer is then used to reconstruct the input using the following transformation:  $\hat{x}=\sigma \left(\tilde{W} y+\tilde{b} \right)$, where  $\tilde{W}: d \rightarrow s$ is a matrix and $\tilde{b} \in R^{s}$.

 Our model used 'tied weights' $\tilde{W}=W^t$, and masked input, as in \cite{vincent2008extracting}, which is equivalent to a dropout layer after the input. This was found experimentally to give the best outcomes (without it, the results degraded significantly). The Adam \cite{kingma2014adam} optimizer was used.
 All the layers except the last used RELU activation. The last layer used the sigmoid activation function $\sigma(x)=\frac{1}{1+e^{-x}}$. 
Batch normalization layers were used \cite{batchnorm} following each layer (except the last) to normalize the activation in the network making it more robust to the difference between the synthetic and clinical paths.





 See Figure \ref{fig:blockdiag} for an illustration of the network.

\label{sec:lossfuncs}
The network was trained using a linear (convex) combination, $L(x,z)$, of the cross entropy loss \cite{vincent2008extracting} and the negative of the DICE coefficient \cite{Milletari2016V-Net:Segmentation}. For a training sample $z$ the cross entropy loss is defined as \[ \addtag CE(x,z) = \sum_{i=1}^n \hat{x}_i \log z_i + (1- \hat{x}_i )  \log (1 - z_i )  \] where summation is over all voxels.
The weighted DICE is defined as : \[ \addtag WDICE(x,y,w_v) = \frac{2 w_v x^T w_v y}{w_v x^T w_v x+ w_v y^T w_v y} \]
Note that the non-weighted DICE is obtained by setting all weights to one. 

\subsubsection*{Spatial Weight Smoothing Regularization (SWR)}
\label{lbl:swr}
In order to reconstruct a realistic atrial volume, we added a \textit{Spatial Weight Smoothing Regularization (SWR)} term to the loss function. 
The loss including SWR  is defined as:  
\[ \addtag L(x,z) + \lambda \sum_{i=1}^n \|  \nabla_{v} W_i \|^2 \]
where $W$ denotes the layer weights and the differentiation is with respect to the spatial dimension $\vec{v}$, that is the position within the volume (for the three spatial axes, $\vec{v}\in (i,j,k)$) in which the input and output reside. $\lambda$ represents the level of regularization.

\begin{figure}[h]
 \centering\includegraphics[scale=0.35]{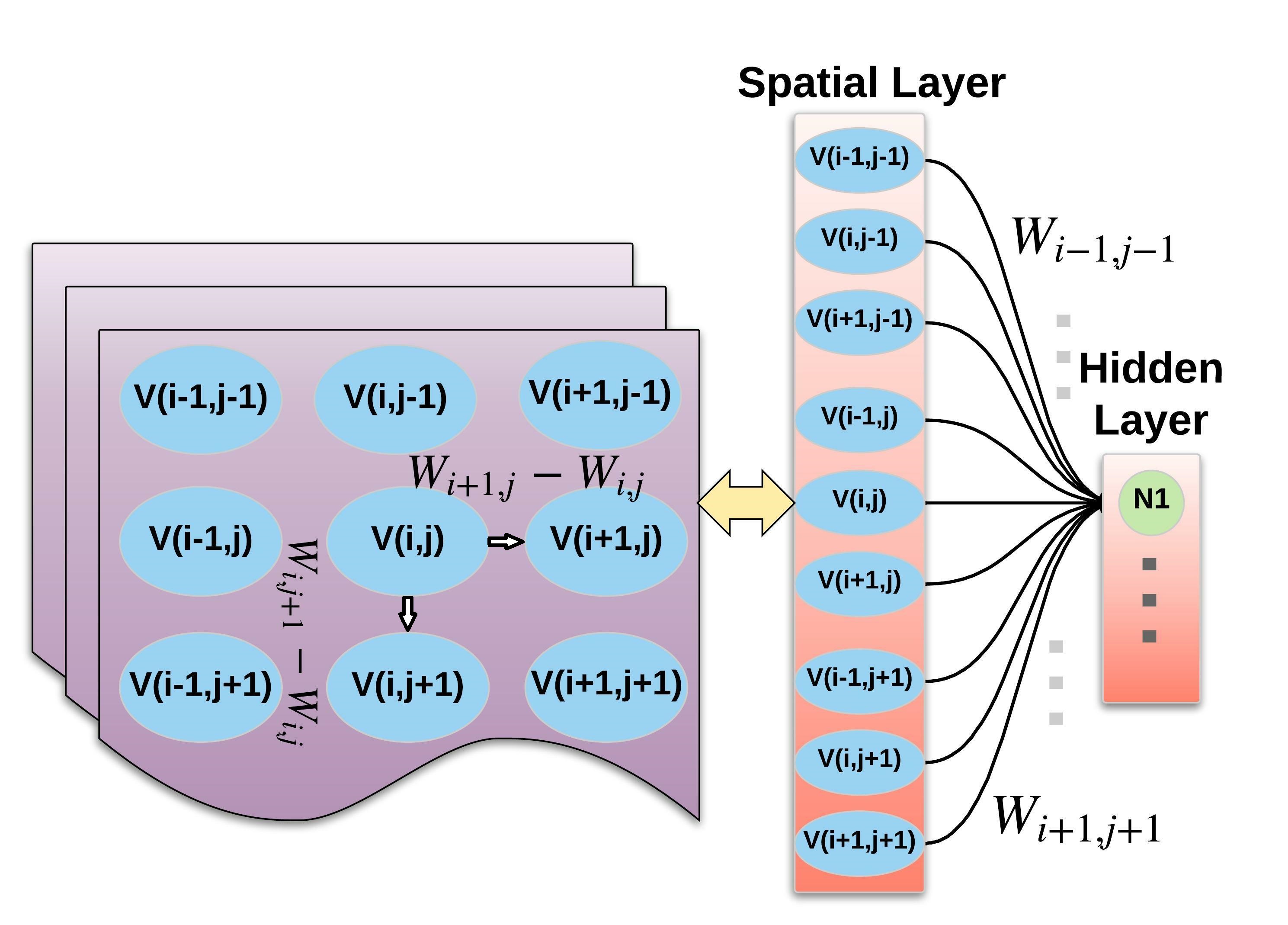}
 \caption{Mapping the volume to the first and last layers. The left part shows a $3$ by $3$ neighborhood in a volume slice and to the right, it's mapping in the network. $V_{i,j}$ represents the voxel in row $i$ and column $j$ in the slice. $W_{i,j}$ is the weight between the spatial layer voxel $V_{i,j}$ and a neuron in the hidden layer. The white arrows indicate the weight derivative between neighbors that is added to the network cost. The slices of the volume are stacked.}
  \label{fig:spatialSmoothing}
 \end{figure}

The SWR loss term is applied to the weights of the input and output layers only,
for which each weight corresponds to a voxel. The spatial derivatives  are computed using finite differences. Figure \ref{fig:spatialSmoothing} depicts the relationship between the voxels in space and the neurons in the layer and shows which weight difference is added to the cost.  

The goal in adding the SWR term  is to have the weights of the first layer that convert volumetric input to output and vice versa (same for the case of tied weights)  be  a smooth  function in $R^3$, as are heart chambers. This obviates the need to use smoothing as a post-processing step. We used a similar concept to the active contours approach \cite{kass1988snakes} where the external energy is the cross entropy loss and the internal energy is SWR.  It is also similar to the interpretation of eigenvectors in PCA. We assume our model learns a blend of shapes that are smooth in nature; each resembles a complete or part of an atrium. The task of the model at inference is to compute which parts to take in order to compute the most probable atrium given the input. 
As a regularization term, this cost helps to prevent over-fitting. 


\begin{figure}
 \centering
  \subfloat[]{\includegraphics[scale = 0.02]{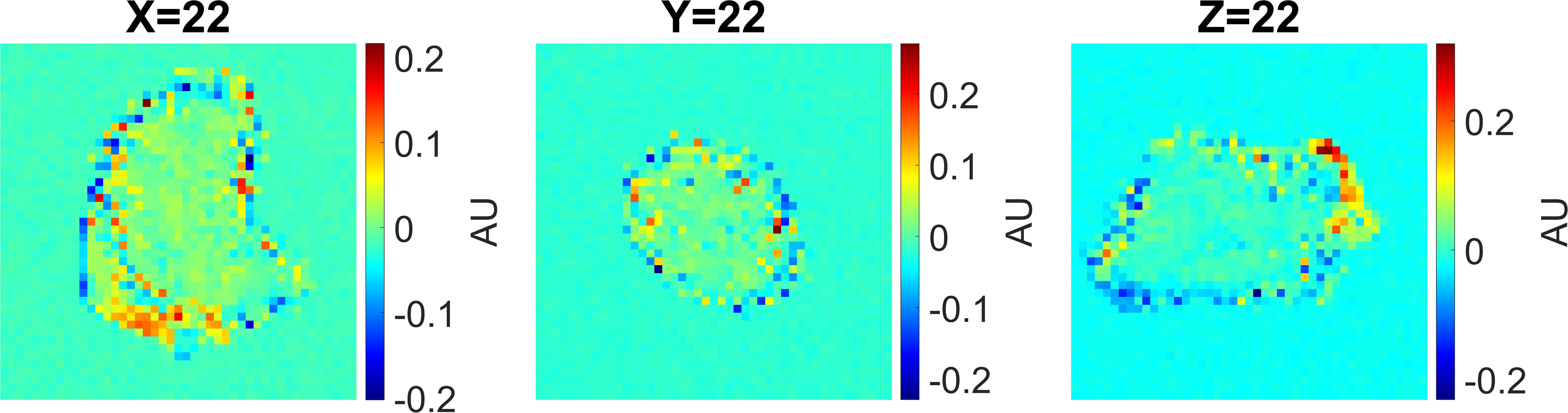}\label{fig:weights_noswr}} 
    \\
     	\subfloat[]{\includegraphics[scale = 0.02]{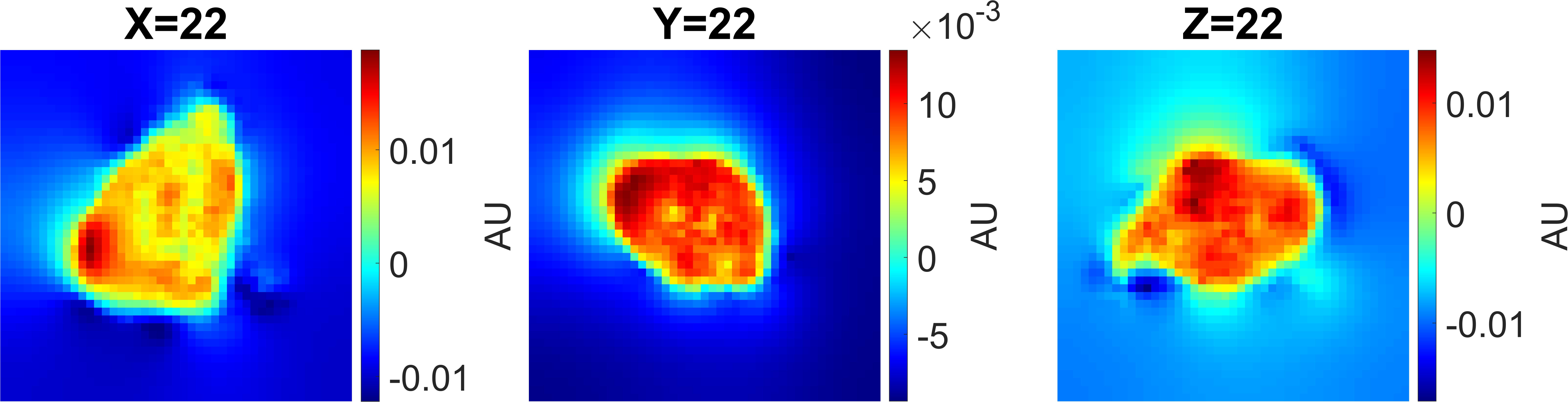}\label{fig:weights_swr}} 
 	\caption{Visualization of the weights from the first layer of \ded \space networks. Slices through the weights volume of the first layer: (a)  shaped as a volume with no SWR. Note the jagged boundary. (b) shaped as a volume with SWR. The resulting atrial boundary is smooth. }
    \label{fig:weights_swrvsnone} 
 \end{figure}

 The effect of the SWR can be seen in Figure \ref{fig:weights_swrvsnone}. We examine a weight from the first layer of two similar networks, one trained with SWR while the other without. The selected weight is reshaped to a $45^3$ voxel volume, maintaining the spatial ordering. In Figure \ref{fig:weights_swrvsnone} we take slices through the three axes of this volume and show these as images. While the network without the SWR learns atria parts with fuzzy boundary, see Figure \ref{fig:weights_noswr}, it is evident that the network with SWR in Figure \ref{fig:weights_swr} learns smooth parts that seem like atria building blocks.
 
\subsubsection*{Boundary Enhancement Mask}
\label{sec:boundaryenhmask}
\begin{figure}
 \centering
  \subfloat[]{\includegraphics[scale = 0.2]{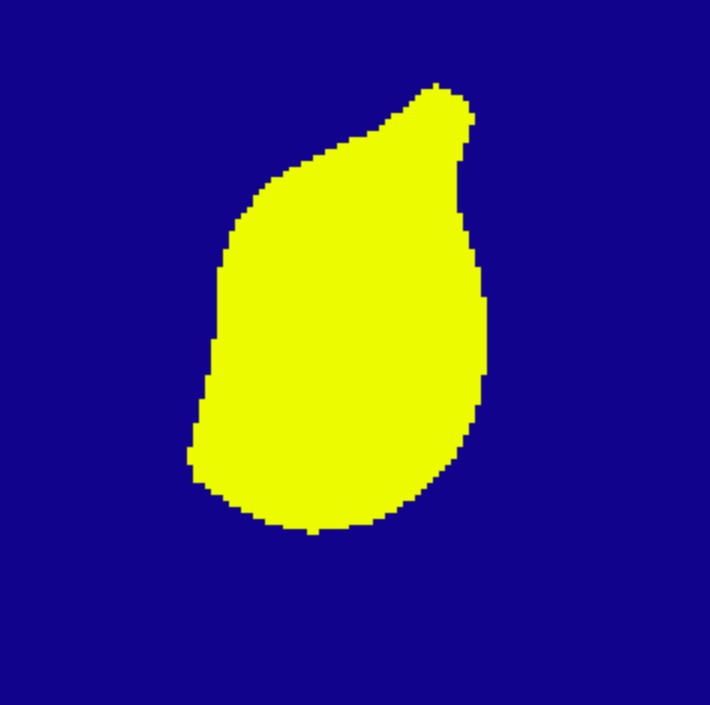}\label{fig:weights_ex_y}} 
    \qquad
     	\subfloat[]{\includegraphics[scale = 0.2]{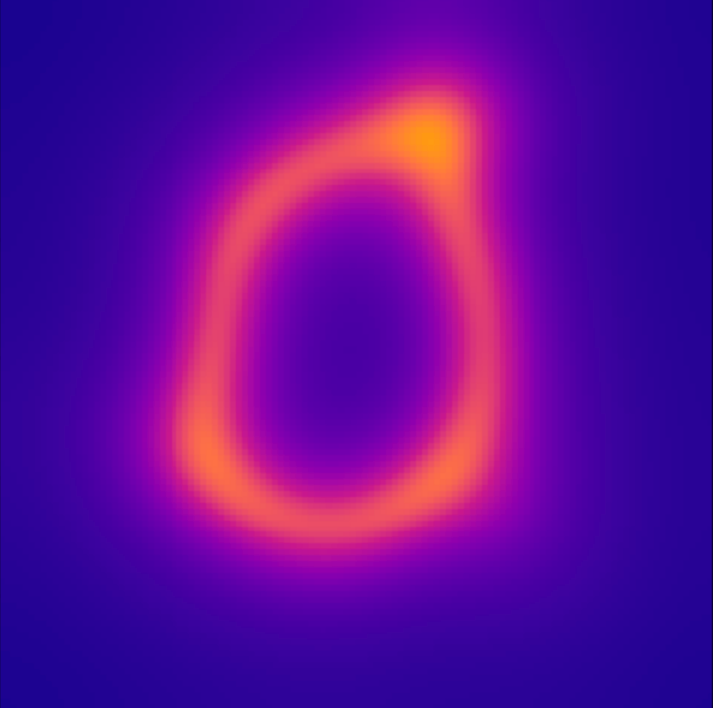}\label{fig:weights_ex_w}} 
 	\caption{Boundary enhancement mask visualization. (a) A slice through the LA Volume, occupied space is in yellow. (b) Corresponding Weight map (arbitrary units), enhancing the LA boundary. High values are indicated by orange.}
    \label{fig:weight_ex} 
 \end{figure}

The most significant area in the volume is the surface boundary.
In order for the model to have larger loss gradients around the surface boundary, we used the weighted DICE cost with a weighting mask $W$. The mask assigns a weight for each voxel such that the majority falls over the boundary, and decreases for voxels further away from it, as seen in Figure \ref{fig:weight_ex}. The weight of a voxel $v$ is given by \[ W = (1+\alpha)/(1+PN(D(v)))  \], where $D(v)$ is the distance from the shape boundary and $PN$ is the probability density function of the normal distribution with zero mean and $\sigma = 1.5$. The $\alpha$ parameter was experimentally set to $14$.

\subsubsection*{\ded \space Parameter Selection}
\label{lbl:chosended}
Subsequent to empirical experimentation, we report the results of the best performing DED variant. During our experimentation, we tested different combinations of depths and widths, see \cite{abaramMiccai2018}. The chosen variant includes two hidden layers with $350$ neurons each. The cross-entropy and (weighted) DICE were combined in the loss function using a ratio of $2:3$.  The variants are named according to the SWR parameter $\lambda$ : "No SWR" with $\lambda=0$, "SWR005" $\lambda=0.05$, "SWR75" $\lambda=75$. "No Aug" net has the same parameters as "SWR005" but with no path augmentation and no boundary enhancement mask.

\subsection{Left Atrial Shape Reconstruction using \vnet \space Network}
\label{lbl:vnet}
The recent success of convolutional neural networks (CNNs) for vision-related tasks motivated us to conduct  experimentation with these networks. Fully convolutional networks provide a classification result for each pixel: The U-Net \cite{Ronneberger2015U-net:Segmentation} for 2D images and the \vnet \space \cite{Milletari2016V-Net:Segmentation} for 3D volumes have shown promising results in many biomedical image processing applications such as breast tissue segmentation in MRI\cite{Dalmis2017UsingVolumes:}, and pancreatic tumor segmentation in CT \cite{Quo2018DeepScans}. Recall that  the \ded \space output, similar to the \vnet, provides a classification label for each voxel in the volume to be either inside or outside the atrium.

The V-Net is a combination of several convolution layers followed by max-pooling, each shrinking the volume by half, after each stage.  This is repeated for four stages. The next steps consist of up-sampling using learned filters  performed over the data in four stages, where residual connections connect the input with information coming from an earlier stage of the same size using concatenation. Batch normalization and dropout are performed to keep the model regularized.
Figure \ref{fig:blockdiag} depicts the architecture of the network and the integration with our system. Similarly to the \ded \space the input is the voxelized catheter path while the output is the LA reconstruction, both represented as occupancy volumes. In addition, network parameters such as the number of filters per layer, the number of down/upsampling layers, and concatenation, are depicted.
The first and last layers have filters with a larger receptive field, which was found to be critical  to the success of the network to provide meaningful results. 


\subsection{Mean Shape as a Baseline Solution}
\label{sec:meanshape}
To determine the effectiveness of any reconstruction algorithm, it needs to be compared to a common solution. In our case, we compared our results to the results of the mean shape solution. 
The mean shape was generated by taking a voxel-wise average over all the ground truth shapes in the training set. 
The mean location of a \PV \space ostia over all the available atria data should converge to that of the mean shape. We defined a base coordinate frame for the reconstruction using the mean shape four \PV \space ostia points.  The input paths from the clinical cases were registered and transformed to have their \PV s ostia match that of the mean shape (in the least squares sense) using rigid point set registration \cite{pointsetregistration}.

\section{Experiments and Results}
\label{sec:expriments}

We next report the experiments and results 
in the reconstruction of the left atrial shape. We start with Synthetic path experimentation in Section \ref{sec:lblsyntexpr} and continue with experiments using human clinical cases, in Section \ref{sec:clinical_cases}.

\subsection{Experiments over Synthetic Paths}
\label{sec:lblsyntexpr}
\subsubsection{Dataset}
These experiments were conducted over $1800$ pairs of paths and the corresponding atria. First, a set of $1800$ LA shapes was created using the LA shape generation described in Section \ref{sec:recscen}. Then, for each LA, a synthetic catheter path was generated using the process described in Section \ref{sec:pathgen}.

\subsubsection{Evaluation Metrics}
\label{lblREvaluationM}
To evaluate the results of the networks, given a ground truth shape, we use two metrics: The first is DICE, which examines the similarity between the resulting volume to the ground truth.
The second metric examines the resulting boundary; It was defined as the set of voxels that separate the interior of the chamber from the exterior. 
To compare the generated and the true boundaries, we  used the average of the distances between all pairs of closest points of the two boundary contours \cite{cha2016urinary} :
\begin{align*} 
\addtag
AVDist(\partial x,\partial y )=0.5  \frac{\sum_{u \in \partial x} \min \lbrace d(u,v):v \in \partial y \rbrace }{\| \partial x \| } + \\  0.5 \frac{\sum_{u \in \partial y} \min \lbrace d(u,v):v \in \partial x \rbrace }{\| \partial y \| }  
\end{align*}

where $d(a,b)$ is the Euclidean distance between voxels $a$ and $b$.


\subsubsection{Results of LA Reconstruction from Synthetic Catheter Paths}

\label{secSynPathExpr}
We ran the set of networks: 'no SWR', 'NO AUG', 'SWR005', 'SWR75', and \vnet \space to produce synthetic data results, i.e. LA reconstructions over the synthesized paths of the test set. 
Table \ref{exprPathRecTable} (a) details the results of the described networks and compares these to the mean-shape solution. Note that for DICE the closer to one the better, while for AVdist the lower the better. 

Evident from the results shown, all the networks were better than the mean shape by approximately $0.7-1mm$ average distance and $0.05$ DICE score. For the DED networks, the networks with SWR outperformed those without. We observe a slight advantage for the 'SWR75' over the rest of the \ded \space networks.
In this testing scenario, the \vnet \space results were strong.
 The run time for our DED networks with two layers over moderate consumer-grade hardware (Nvidia GTX2080) was less than 0.9ms for 20 samples.

\begin{table*}
\centering
\caption{LA Reconstruction from synthetic paths inputs and clinical cases, comparing DED, \vnet , and Mean shape results. }
\label{exprPathRecTable}
\begin{tabular}{l|l|llll|l|l|} 
\cline{2-8}
\multirow{2}{*}{}                         & \multirow{2}{*}{}            & \multicolumn{4}{l|}{\textbf{Dense Encoder Decoder (DED) }}              & \multirow{2}{*}{\textbf{VNet}} & \multirow{2}{*}{\textbf{Mean Shape}}  \\ 
\cline{3-6}
                                          &                              & \textbf{No Aug} & \textbf{No SWR} & \textbf{0.05 SWR} & \textbf{75 SWR} &                                &                                       \\ 
\cline{2-8}
\multirow{2}{*}{\textbf{(a) Synthetic Paths}} & \textbf{DICE}                & \textit{0.949} & 0.942          & 0.946            & 0.949           & \textbf{0.956}                 & 0.895                                \\ 
\cline{2-8}
                                          & \textbf{AVDist(mm)}          & 1.331           & 1.45            & 1.388            & \textit{1.32} & \textbf{1.204}                & 2.261                                 \\ 
\hhline{~=======|}
\multirow{3}{*}{\textbf{(b) Clinical Cases}}                   & 
\textbf{\textbf{Mean Distance(mm)}} & 5.082           & N/A             & 5.137             & \textbf{5.015}  & 5.256                          & 5.79                                  \\ 

& \textbf{\textbf{Standard Deviation}} & 1.782           & N/A             & 1.863             &1.058  & 1.032                          & 2.995 \\

& \textbf{\textbf{p-value ($<$)}} & 0.076           & N/A             & \textbf{0.047}             &0.124  & 0.249                          & N/A \\
\hhline{~-------|}
\end{tabular}
\end{table*}

\subsection{Experiments in Human Clinical Cases}
\label{sec:clinical_cases}
The networks with the best-performing parameter sets over the synthetic data were used to examine the feasibility of the proposed methods on actual human clinical cases. 
\subsubsection{Data Acquisition and Network Input Generation}
\begin{figure}
  \centering
  \subfloat[Focal Catheter Point Cloud Path]{\includegraphics[scale = 0.07]{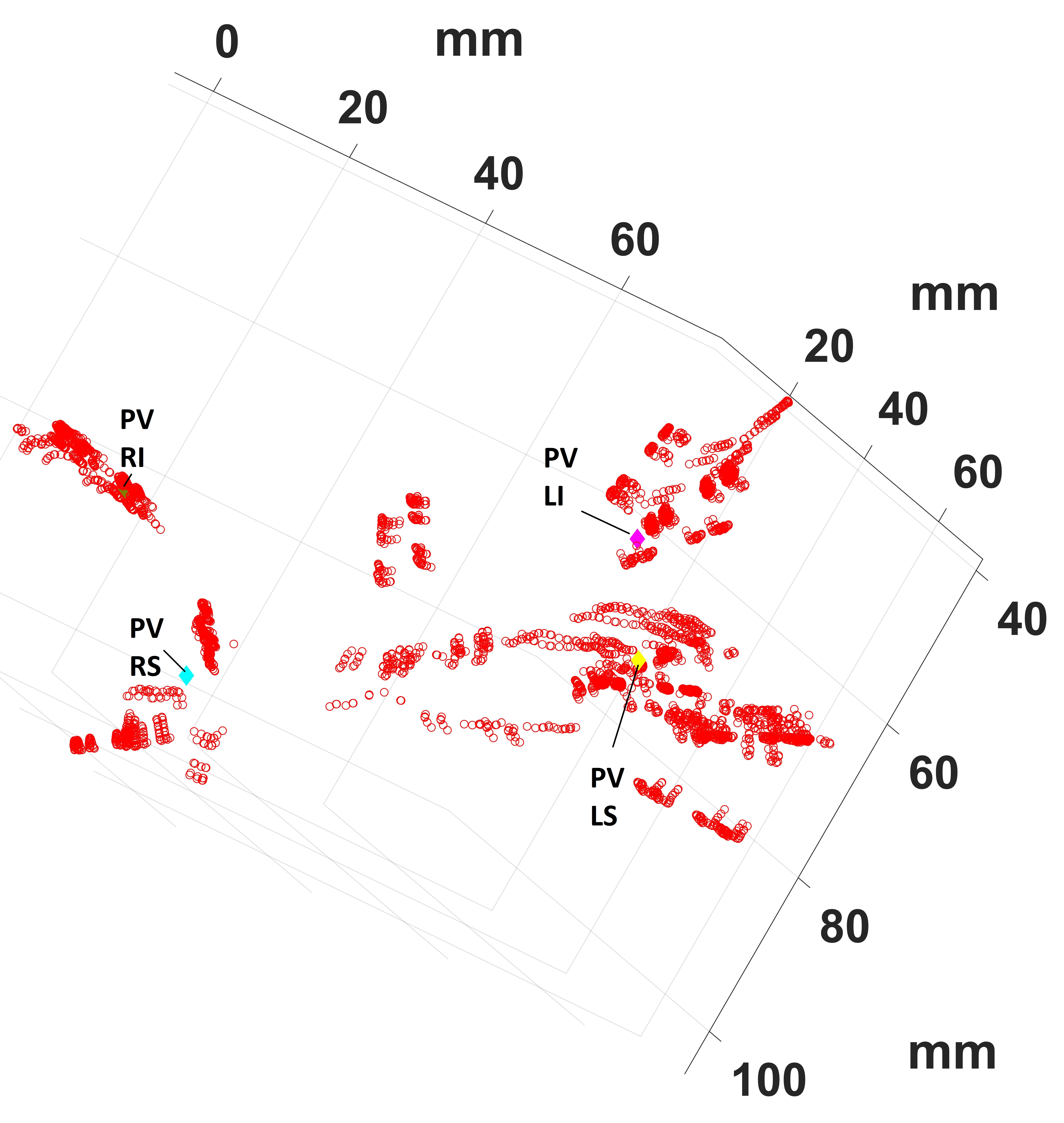} \label{fig:EEInputA}} \

  \subfloat[Focal Catheter Voxelized Path]{\includegraphics[scale = 0.16]{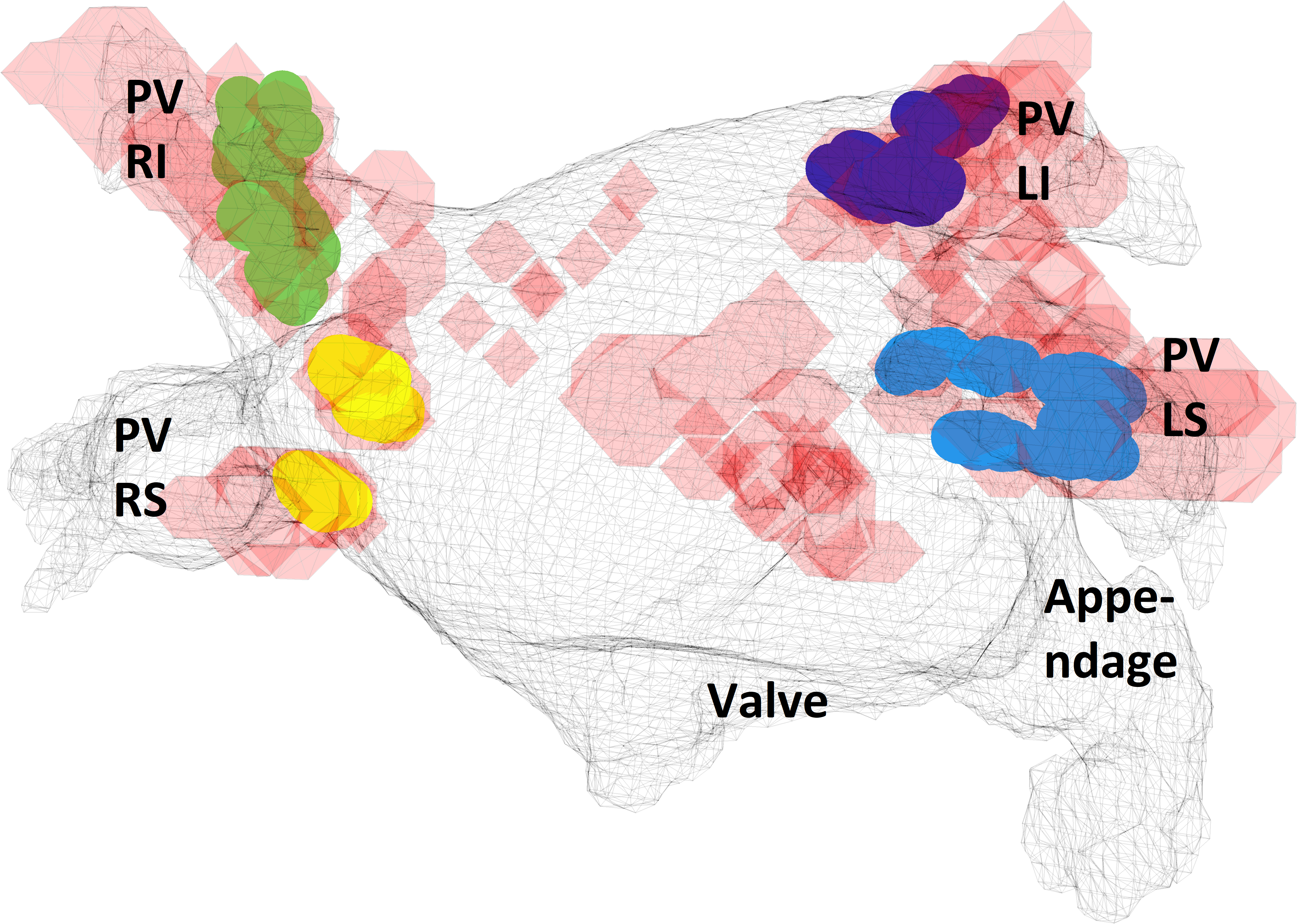} \label{fig:EEInputB}} 
  \subfloat[Lasso Catheter Voxelized Path]{\includegraphics[scale = 0.0525]{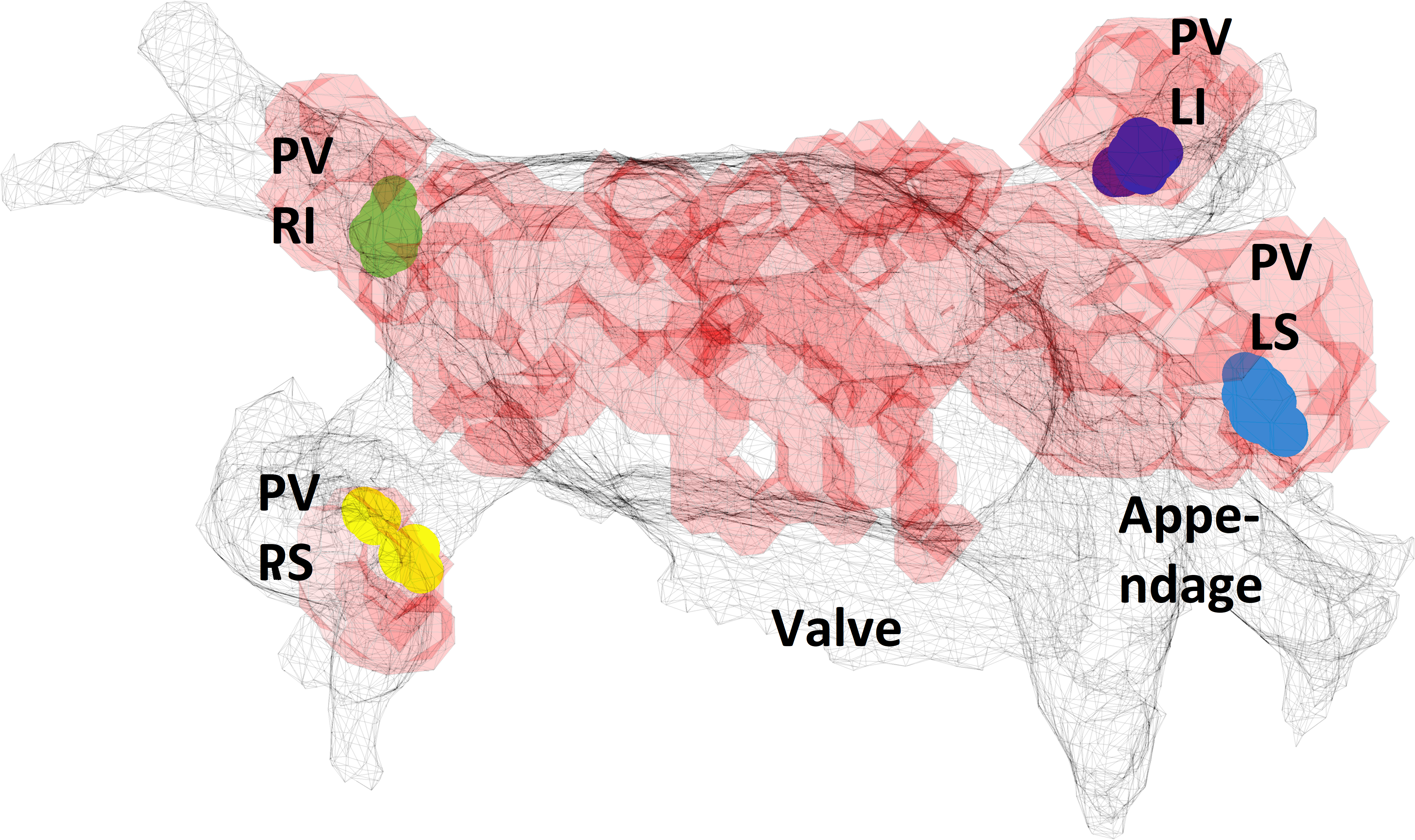} \label{fig:EEInputC}}\
  \subfloat[Point cloud of a focal catheter registered with the blue path of the mean shape.]{\includegraphics[scale = 0.06]{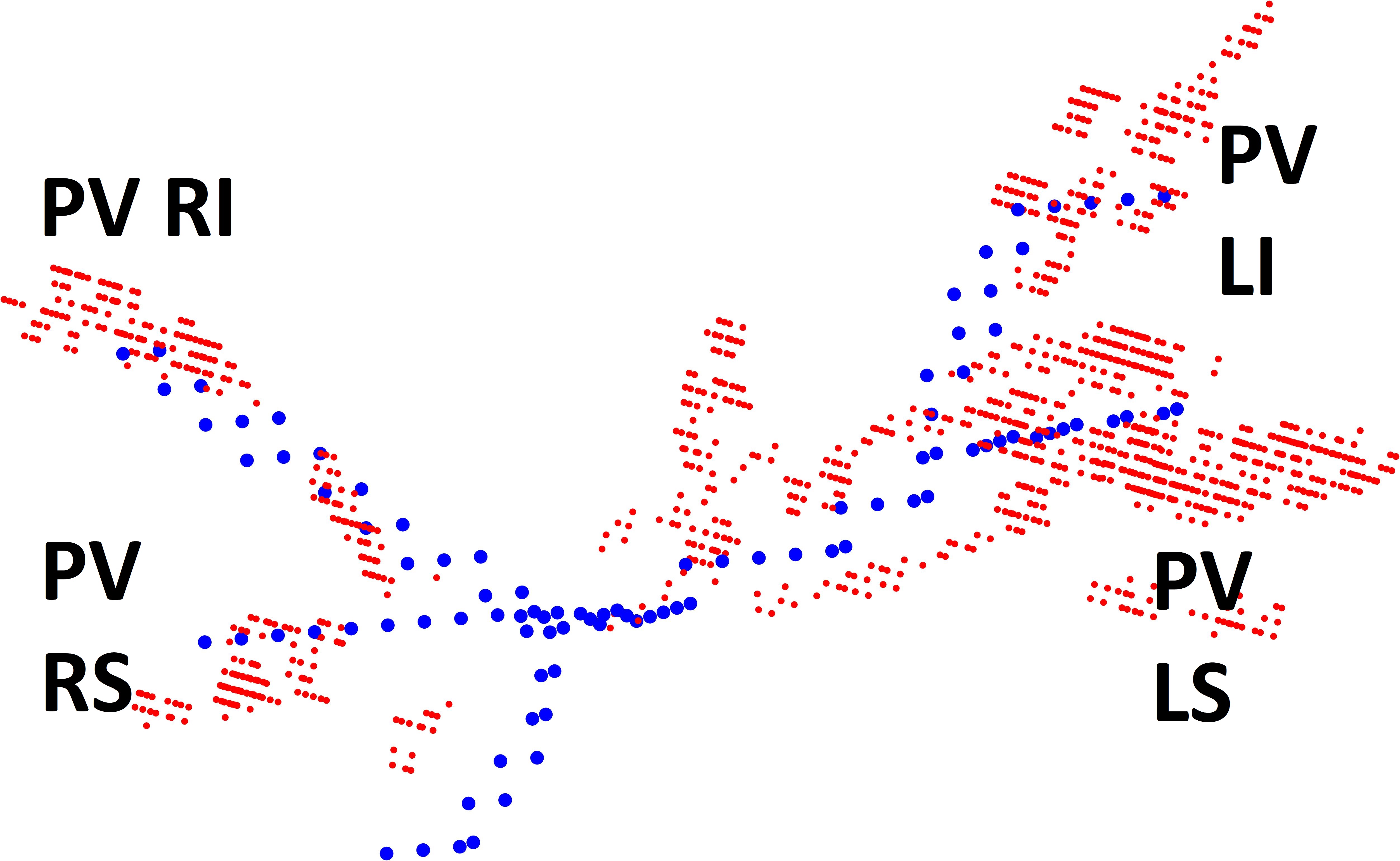} \label{fig:EEInputD}} 
 \caption{Input examples for the clinical cases. The acquired path is in red, the synthetic template path in blue and the CT is in gray. Tagged \PV \space points are color coded.}
  \label{fig:EEInputOutout}
 \end{figure}

Our clinical dataset was composed of $80$ cases, for which different catheters were used. Twenty-six of the cases had a properly registered mesh resulting from a CT segmentation. At the beginning of a clinical procedure, the initial bearing path that starts from the septum, and connects the \PV s was acquired, with an acquisition time of fewer than three minutes. The physician tagged the points belonging to each \PV \space ostia. Figure \ref{fig:EEInputA} illustrates the input point cloud (path) with tagged \PV s centroids. Figures \ref{fig:EEInputB} and \ref{fig:EEInputC} show the input volumes in red, with colored tagged points for each \PV \space (PVRS  yellow, PVRI green, PVLI blue, PVLS light blue) and the registered CT. Note that though the path performed in the clinic traversed similar locations, it differed greatly from the generated synthetic paths. Moreover, when the catheter exceeded a predefined velocity, position acquisition is suspended. This created discontinuities in the acquired path. Other differences can be attributed to deviations in the declared protocol from our definitions (visiting other locations, not traversing all locations for our path), the use of multiple arm catheters, different catheter maneuvers, and so on.

The input to the network was generated by implementing the following steps. First, we need to perform coordinate systems alignment between the acquired path coordinate system and the coordinate system of the networks (which corresponds to that of the mean shape, see Section \ref{sec:meanshape}).  We find each \PV \space ostia of the acquired point cloud by taking the mean of respective \PV \space tagged points.
To find the transformation between the two coordinate systems, we applied a rigid point set registration, \cite{pointsetregistration} matching the \PV \space ostia of the acquired point cloud to the four \PV \space ostia points of the mean shape, see Figure \ref{fig:EEInputD}. This transformation is then applied to the acquired point cloud.
In the following step, we convert the transformed acquired point cloud to an occupancy volume. We sampled the  point cloud in a $45^3$ voxel volume where each voxel indicated the presence (yes/no) of a point (or several) of the input cloud within the voxel ($2.666mm^3$ per voxel). This volume was the expected input to our network. We note that the input point cloud may differ with catheter type. Figure \ref{fig:EEInputB} depicts the input volume for a focal catheter while Figure \ref{fig:EEInputC} shows a volume resulting from using a round \lassocath, where rings of voxels are acquired together.

The network output volume was converted to a triangular mesh by using the Marching cubes algorithm \cite{marchingcubes} implemented in Matlab. The mesh was defined as the iso-surface at a value of $0.5$ of the output volume.

We analyzed the performance of the reconstructions using two evaluation methods, the first using contact points while the second compares to ground truth CTs.
\subsubsection{Evaluation by Contact Points}
Some of the cases used a force sensing catheter (\sfcath) to acquire points that lay on the surface; namely, contact points with a contact force between $5g$ and $15g$, when respiratory gated (end-expirium). These are of interest to physicians since their position is accurate (less than $1mm$ error) and are usually located near ablation regions and important anatomical landmarks. 
 For this evaluation, we chose cases that use a focal catheter. We included cases in which the acquired point cloud covered most of the synthetic path parts. In addition, there had to be enough ground-truth points located in areas of interest such as the \PV s. A total of nine cases were suitable.

The reconstruction accuracy was measured as the mean of the distances of the ground truth points from the reconstructed shape. In particular, we measured the distances from each ground truth point to the nearest mesh vertex. 

\begin{figure*}
 \centering
  \subfloat[Sample A]{\includegraphics[scale=0.12]{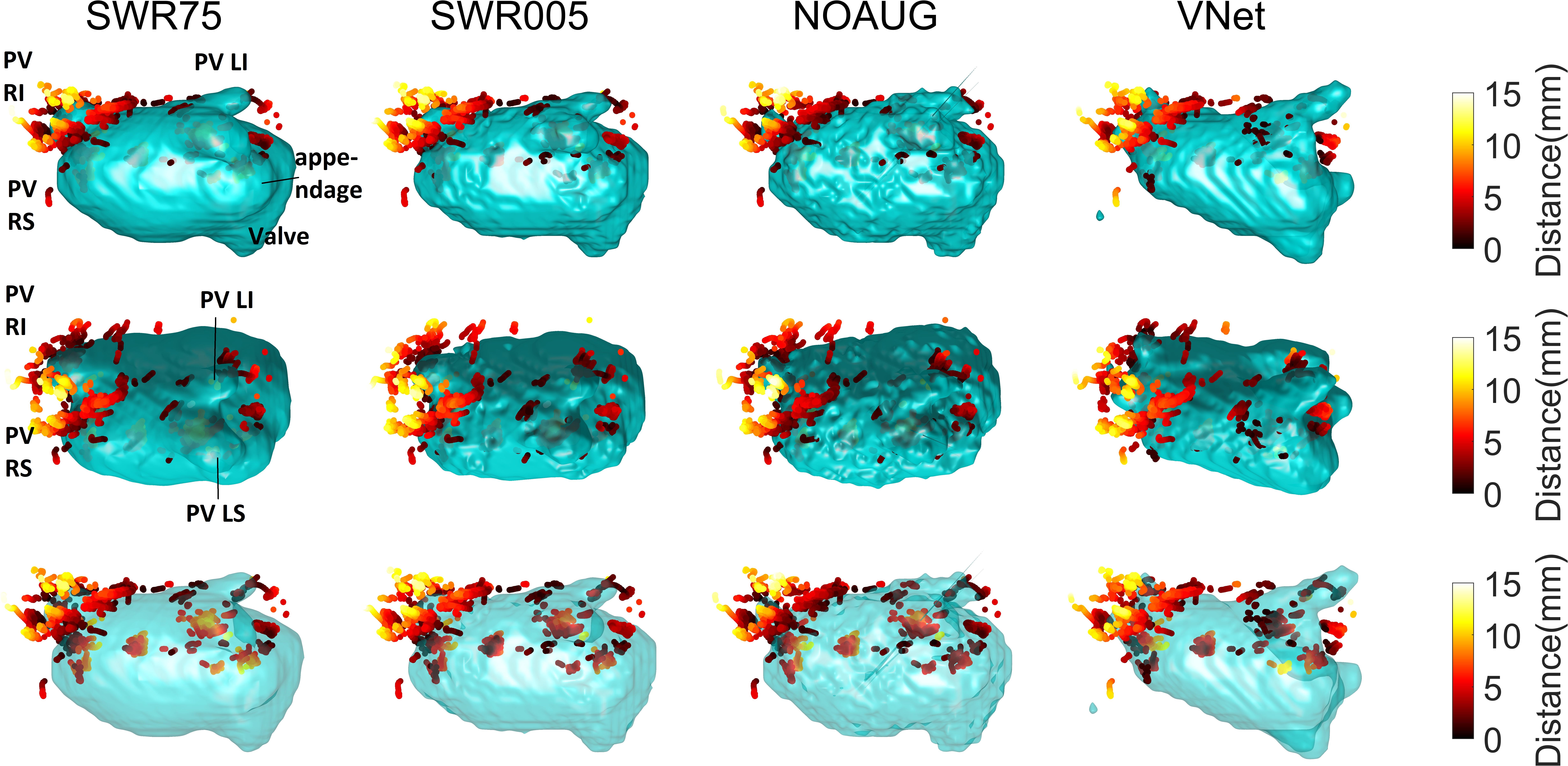}\label{EERecOne}} \\
  \subfloat[Sample B]{\includegraphics[scale=0.12]{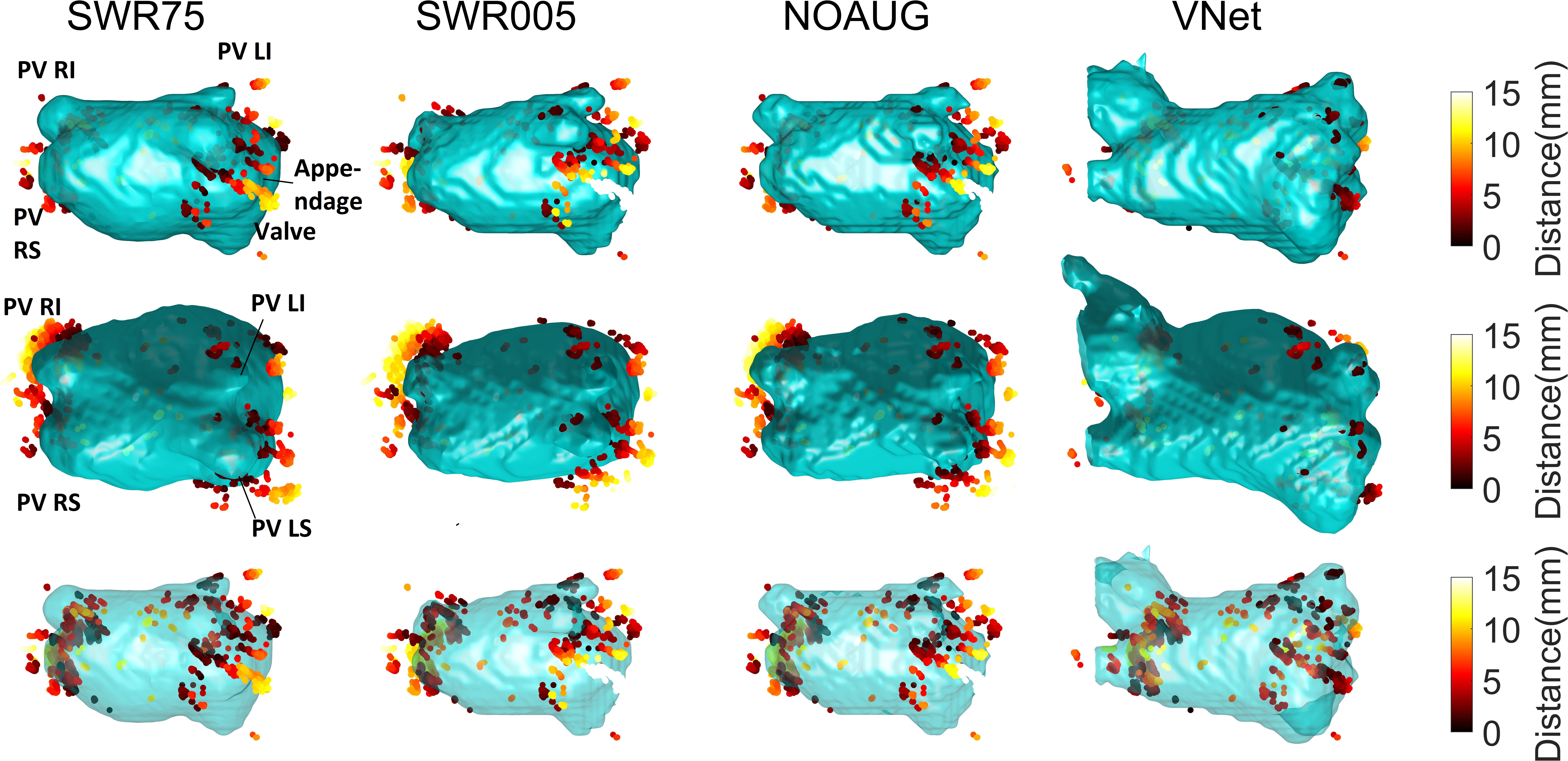}\label{EERecTwo}}
 \caption{Reconstruction result for 'SWR75','SWR005','NOAUG', and \vnet \space networks. On the top is a side view, middle is a top-down of the \PV s and the bottom is a transparent reconstruction showing all ground truth points, color-coded by distance to reconstruction. }
  \label{fig:RecCases}
 \end{figure*}

We present the results for the three networks defined in Section \ref{lbl:chosended}, the \vnet, and the mean shape.
The reconstruction accuracy results are compared in Table \ref{exprPathRecTable} (b). It shows that the errors were highly comparable, although \vnet \space lagged behind. All the results were highly better than the mean shape; however, due to the variance, the differences in means were significant only for 'SWR005' (paired t-test, $pvalue<0.05$), while improving the mean by $0.84mm$. 
We omit the results of \ded \space without SWR since it produced invalid results for some of the samples (jagged disconnected results with no interpretable boundaries).     
Figure \ref{fig:RecCases} show the resulting surfaces of the tested networks for two cases, overlaid with ground truth points colored by distance to reconstruction.
Two views are shown, while the third is a transparent view showing all the ground truth points, colored by distance to the reconstruction.
 'SWR005' reconstructions appear smoother than 'NOAUG' reconstructions with more apparent anatomy, thus only 'SWR005' is kept for the next evaluation. 'SWR75' produced better smooth-looking shapes at the expense of making the ridge between the left PVs and the appendage less distinguishable.

\subsubsection{Evaluation with Ground Truth CT }


We gathered a total of $26$ cases with a CT that was properly registered to the acquired point cloud representing the catheter traversed location. This enabled us to use the same coordinate system alignment between the point cloud and the network coordinate system found earlier to bring the CT to align with the network reconstruction result. 

\begin{figure*}

 \centering
  \includegraphics[scale = 0.55]{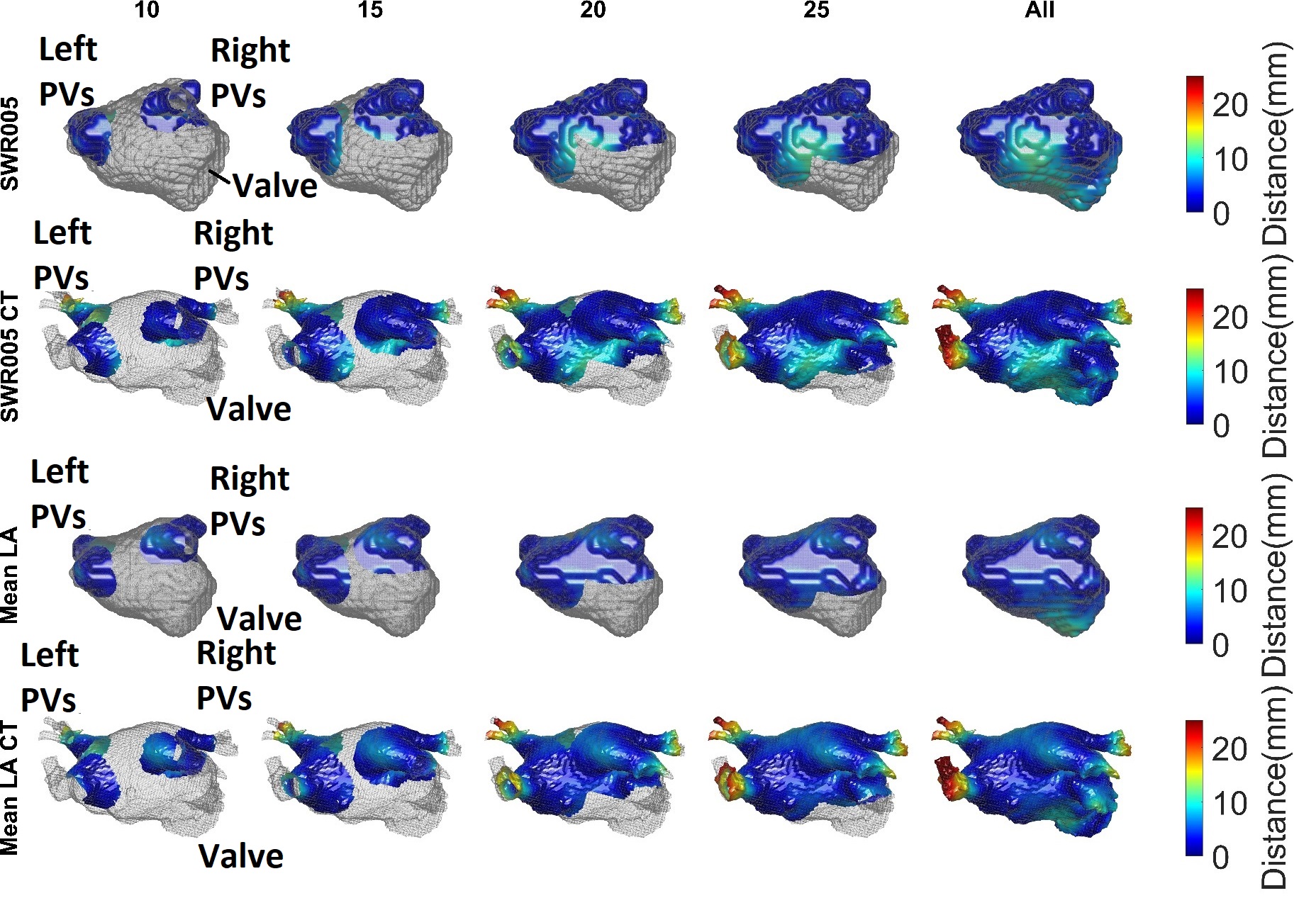}
 
    \caption{Comparison of SWR005 and mean shape surface to surface distance for several radii (mm). Each row depicts the same view with different radii, while 'All' stands for unbounded radius. The first row for each model shows the reconstruction to CT distances while the second row is the CT to reconstruction distances. }
\label{fig:eect_radii_mean_shape}
\end{figure*}

Next, we compared the extracted CT surfaces to the reconstructed mesh using the surface to surface distances. Since this metric is largely affected by the surface area of the measurement which is irrelevant in large areas of the LA body we focus on clinically relevant regions. This was done by considering the surface to surface distance within some radius of the physician-tagged points. In this evaluation, the surface to surface distance was computed as follows: 
Given a vertex on the first mesh, we took the nearest vertex on the second mesh and kept the distance. The total distance is the average overall examined vertices. For symmetry, we averaged both the distance from the reconstruction to the CT and vice versa. To examine regions important for the clinic, we only took vertices that were within defined radios of tagged \PV \space points. Figure \ref{fig:eect_radii_mean_shape} illustrates the measured regions for possible choice of radii. It is evident that the radius of $15mm$ captures the \PV s ostia surroundings well.

The quantitative results for all $26$ CT cases, comparing the surface-to-surface distances for 'SWR005', 'SWR75', \vnet, and the mean shape in various radii are summarized in Table \ref{exprPathRecCTTable}, for all cases. The reported p-value indicates the significance of improvement vs. the mean shape (paired t-test, one-tailed). Quantitatively, 'SWR75', 'SWR005' and \vnet \space improved the distance metric by $0.3-0.45mm$ (statistically significant) over the mean shape results for radii between $15-25mm$.

\begin{table*}[]
\centering
\caption{Surface to surface distances (mean and standard deviation) comparing LA reconstructions to ground truth CT over $26$ clinical cases. Results are shown for \ded, \vnet, and Mean shape, in four different radii, and unbounded distance. The p-value tests for significant differences between the network and the mean shape. P-value under the significance level of $0.05$ is in \it{italics}. }
\label{exprPathRecCTTable}
\resizebox{0.7\textwidth}{!}{
\begin{tabular}{|ll|l|l|l|l|l|}
\hline
\multicolumn{2}{|l|}{\textbf{Distance from Interest Points}}                                  & \textbf{Unbounded} & \textbf{10mm}  & \textbf{15mm}  & \textbf{20mm}  & \textbf{25mm}  \\ \hline
\multicolumn{1}{|l|}{\multirow{3}{*}{\textbf{DED 0.05 SWR}}} & \textbf{Mean}                  & 6.057              & \textbf{4.307} & 4.697          & 5.083          & 5.363          \\
\multicolumn{1}{|l|}{}                                       & \textbf{Std}                   & 0.858             & 0.888        & 0.832         & 0.868         & 0.921         \\
\multicolumn{1}{|l|}{}                                       & \textbf{p-value (\textless{})} & 0.055              & \it{0.00054}        & \it{0.00039}        & \it{0.00058}        & \it{0.0023}         \\ \hline
\multicolumn{1}{|l|}{\multirow{3}{*}{\textbf{DED 75 SWR}}}   & \textbf{Mean}                  & 6.149              & 4.327          & \textbf{4.678} & 5.093          & 5.357          \\
\multicolumn{1}{|l|}{}                                       & \textbf{Std}                   & 0.979             & 0.803         & 0.712         & 0.767         & 0.837         \\
\multicolumn{1}{|l|}{}                                       & \textbf{p-value (\textless{})} & 0.24               & \it{0.0091}         & \it{0.0045}         & \it{0.009}          & \it{0.016}          \\ \hline
\multicolumn{1}{|l|}{\multirow{3}{*}{\textbf{VNET}}}         & \textbf{Mean}                  & \textbf{5.962}     & 4.441          & \textbf{4.678} & \textbf{4.984} & \textbf{5.216} \\
\multicolumn{1}{|l|}{}                                       & \textbf{Std}                   & 0.913             & 0.75         & 0.708         & 0.651         & 0.662         \\
\multicolumn{1}{|l|}{}                                       & \textbf{p-value (\textless{})} & 0.068              & 0.14           & \it{0.045}          & \it{0.025}          & \it{0.022}          \\ \hline
\multicolumn{1}{|l|}{\multirow{2}{*}{\textbf{Mean Atrium}}}  & \textbf{Mean}                  & 6.241              & 4.735          & 5.103          & 5.458          & 5.686          \\
\multicolumn{1}{|l|}{}                                       & \textbf{Std}                   & 1.225              & 1.311          & 1.195          & 1.19           & 1.244          \\ \hline
\end{tabular}
}
\end{table*}

\begin{figure*}

 \centering
  \subfloat[]{\includegraphics[scale = 0.25]{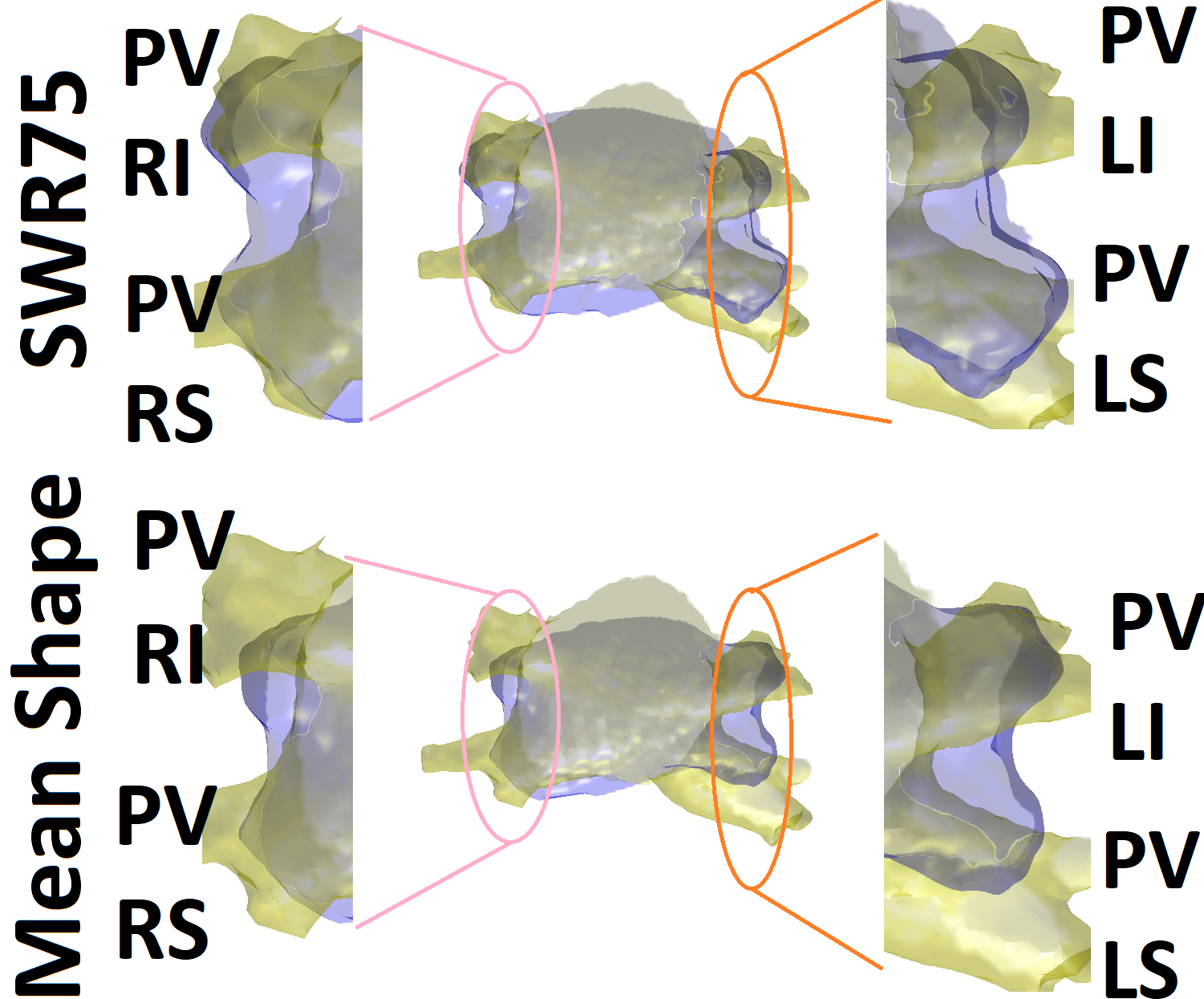}} \qquad
  \subfloat[]{\includegraphics[scale = 0.22]{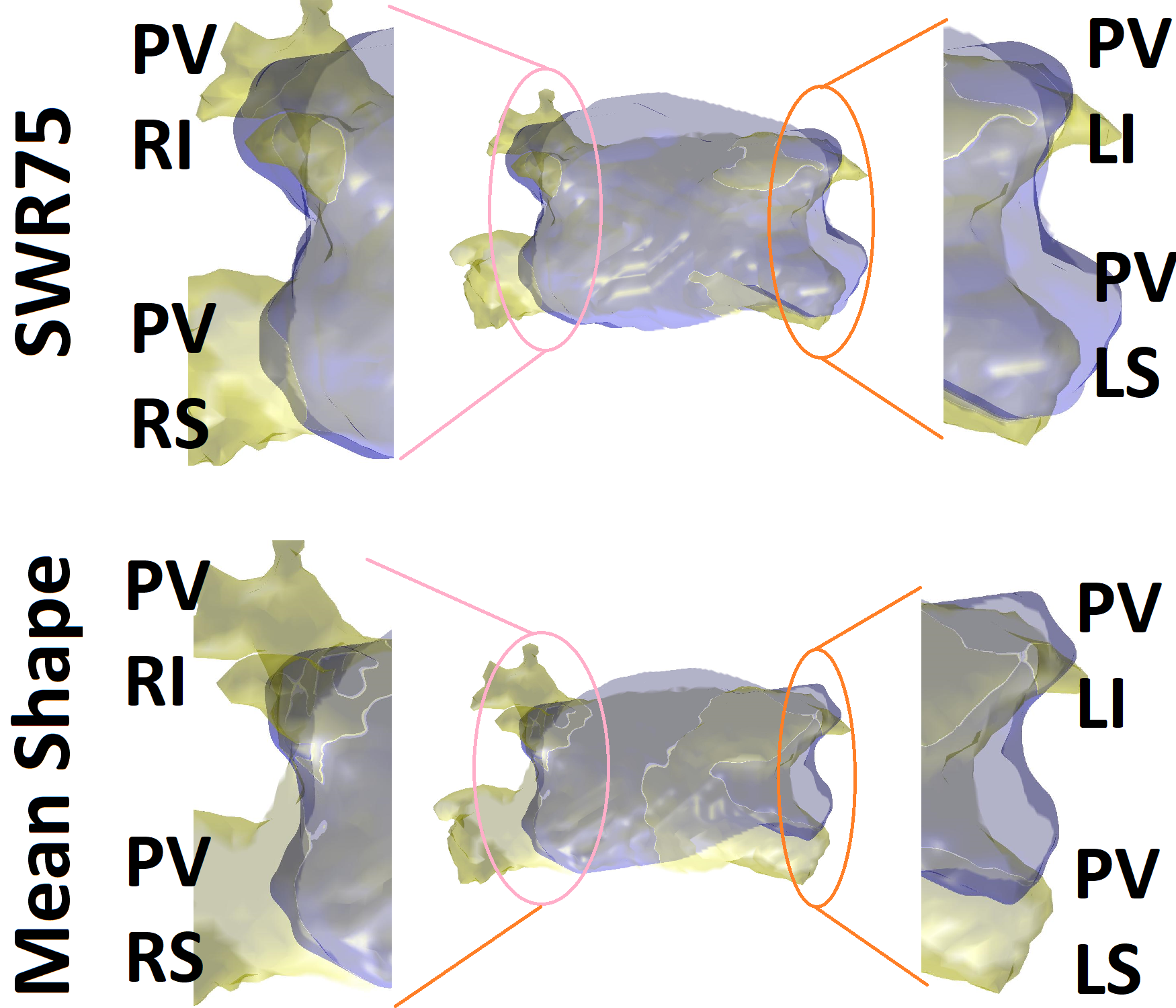}}
    \caption{Comparison of reconstruction results focusing on the location and orientation of the PVs, for 2 sample cases:  SWR75 reconstruction (top)  mean shape reconstruction (bottom).  CT ground truth is shown in yellow, and the reconstruction  results in blue. The visualization is clipped such that only the PV areas are shown. We note the better overlap of the blue and yellow regions in the SWR case.} 
\label{fig:eect_swr_macmp_a}
\end{figure*}

\begin{figure*}

 \centering
  \includegraphics[scale = 0.58]{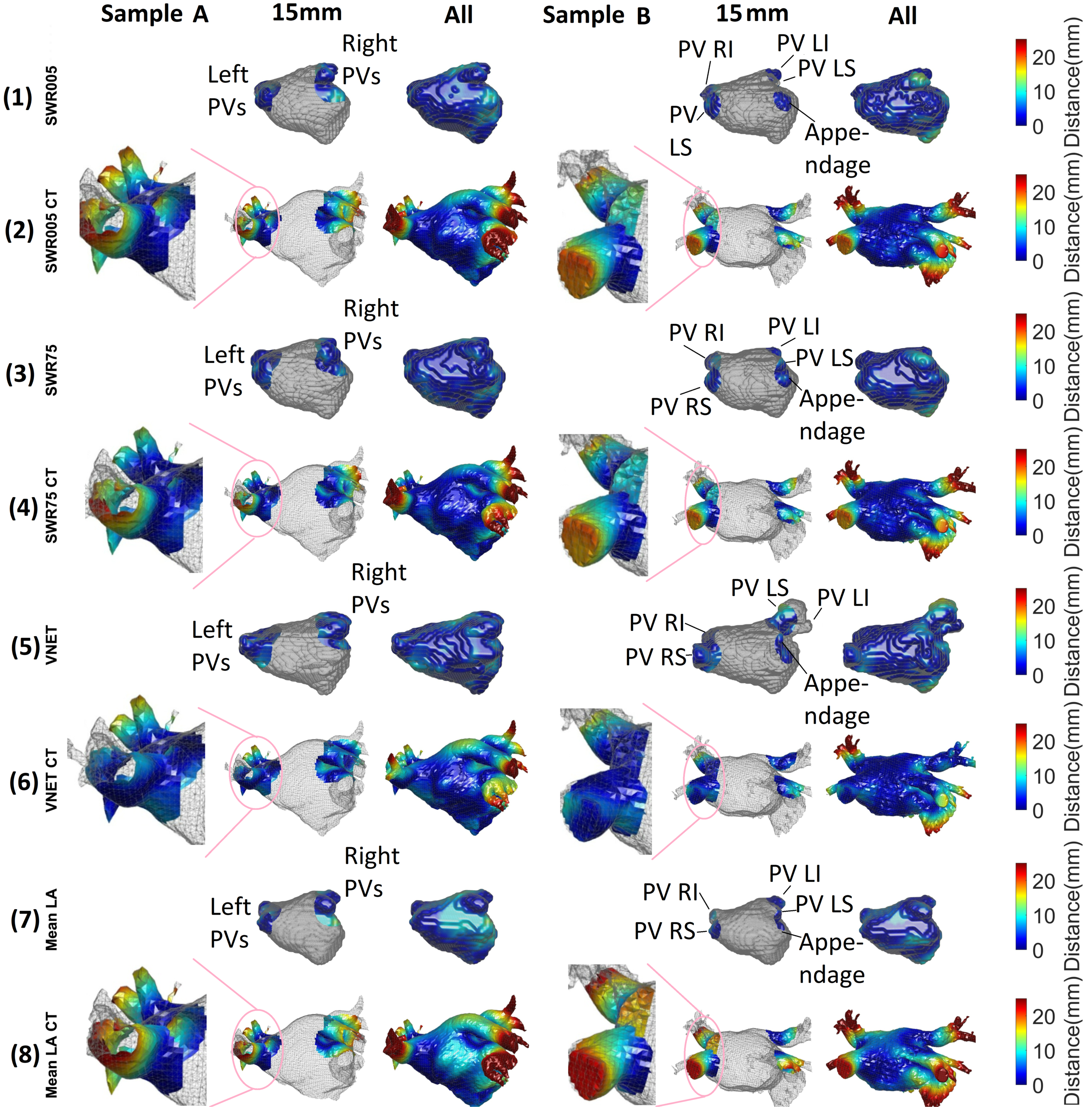}
 
    \caption{Visualization of the reconstructions in the clinical CT cases for the SWR005, SWR75, \vnet \space, and the mean shape. The surface to surface errors between the reconstruction and the CT, for $15mm$ radius and unbounded radius, are shown. Each row is a similar view, where the first shows the reconstructions along with distances to the CT, while the second is the CT with distances to the reconstruction. Error with reconstructed \PV s mostly affects the CT to reconstruction distance and relevant sections are zoomed in. By comparing these regions we see the improvement of the different reconstructions over the mean shape solution.  }
\label{fig:eect_swr_macmp_all}
\end{figure*}

\begin{figure*}

 \centering
  \subfloat[V-Net Sample A]{\includegraphics[scale = 0.221]{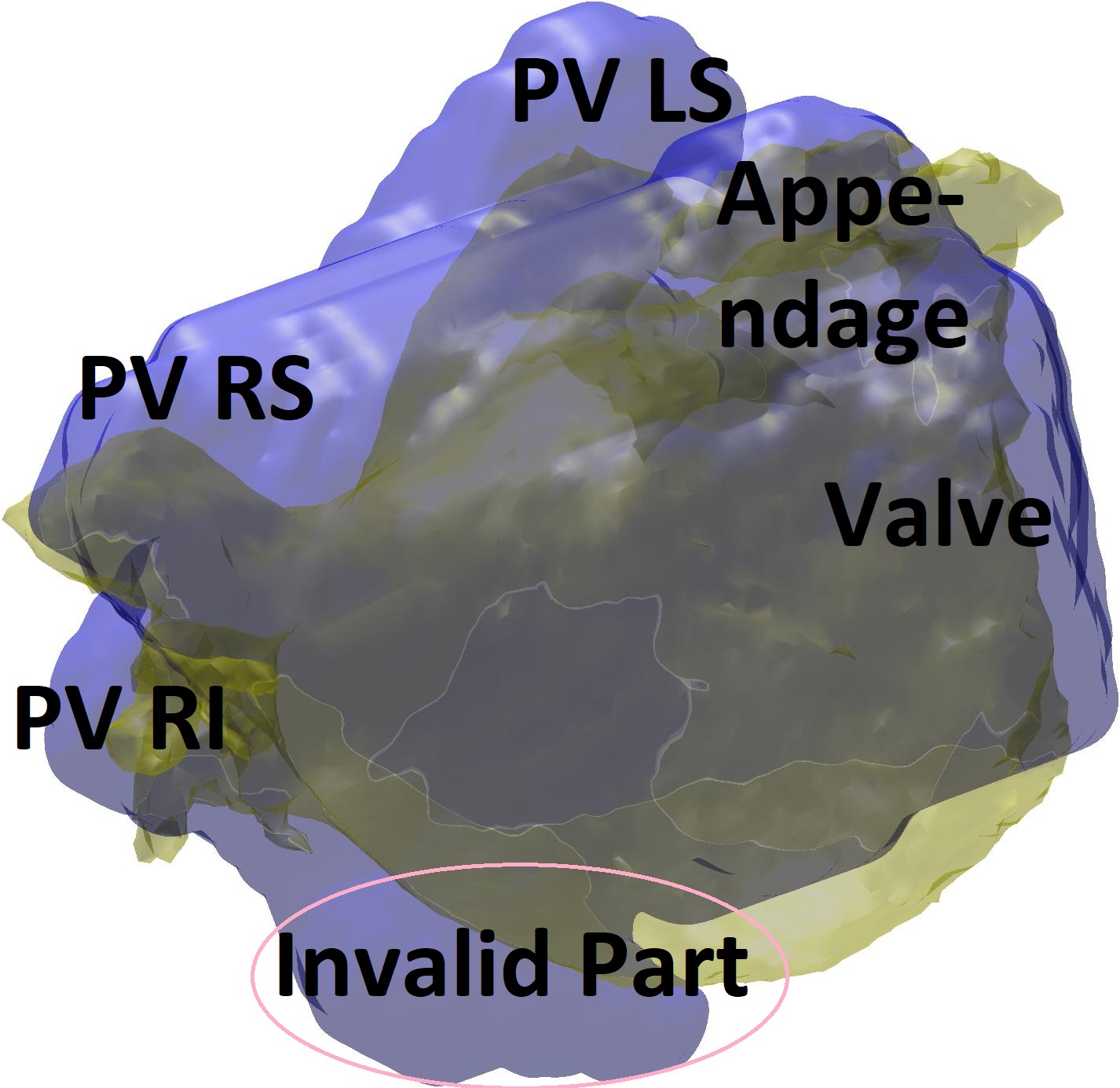}} \qquad
  \subfloat[V-Net Sample B]{\includegraphics[scale = 0.221]{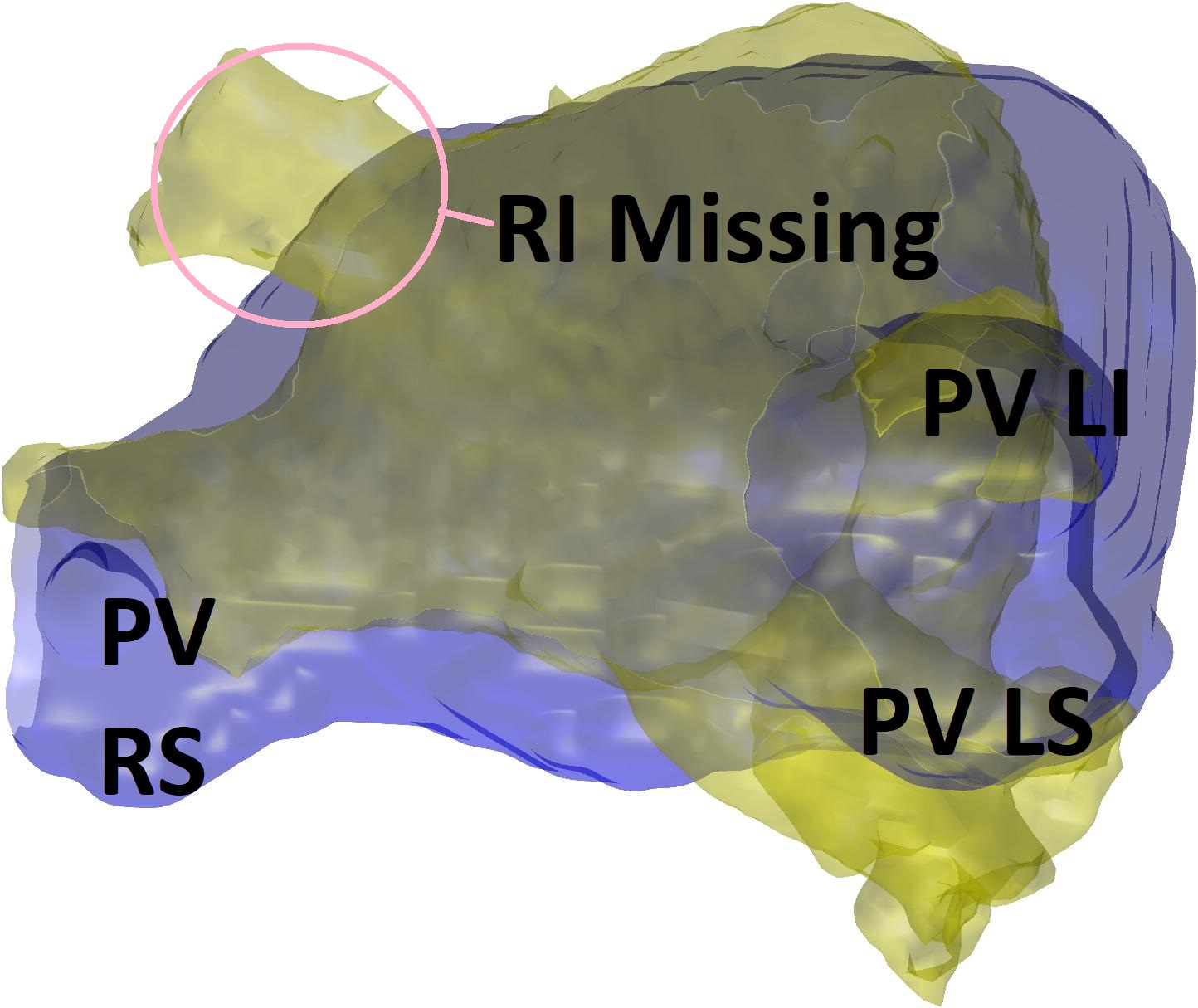}} \qquad
  \subfloat[V-Net Sample C]{\includegraphics[scale = 0.15]{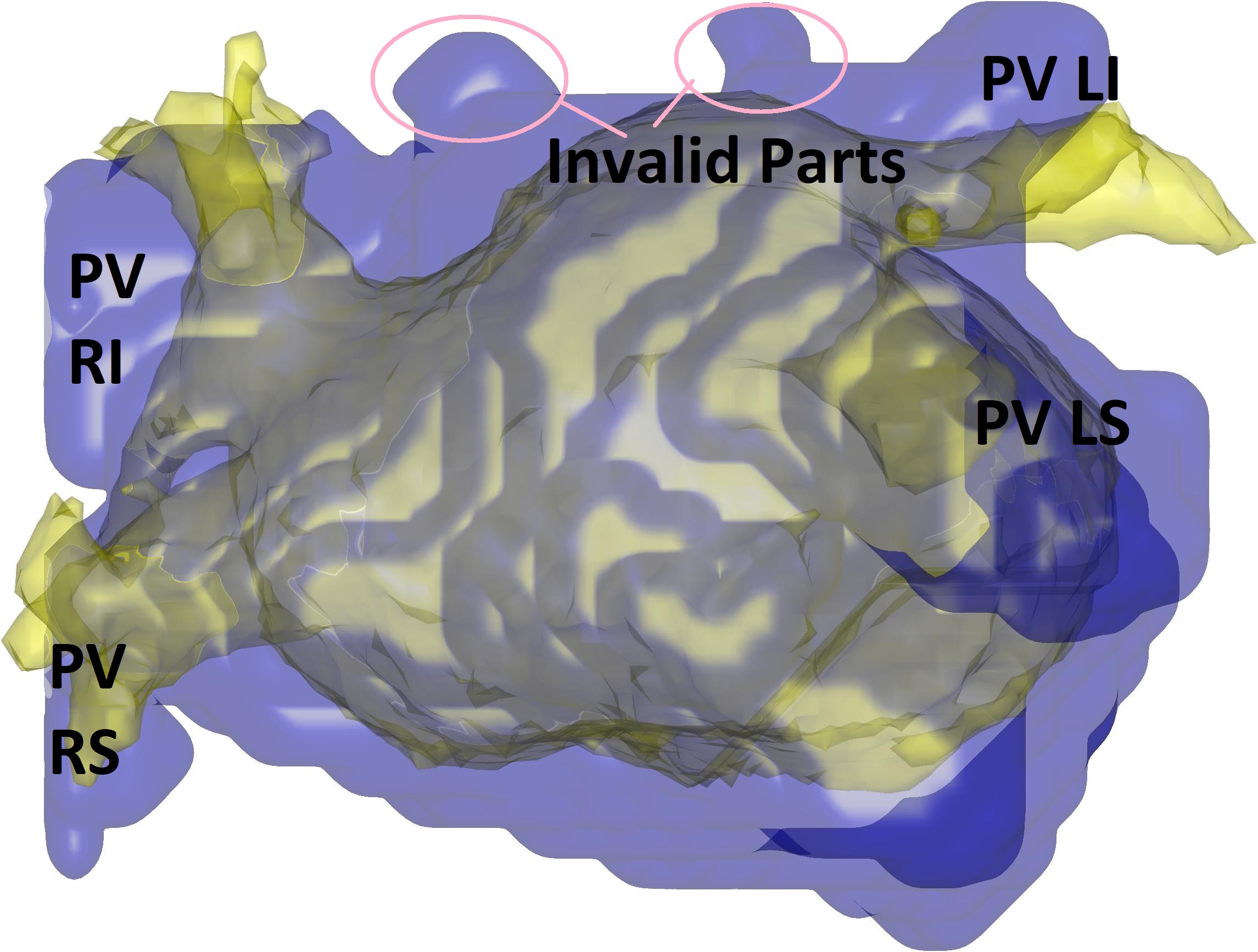}}
    \caption{ Invalid anatomy in V-Net reconstructions. The CT in yellow, reconstruction in blue, invalid anatomy is circled in pink.}
\label{fig:eect_vnet_iregular}
\end{figure*}

\begin{figure}

 \centering
  \subfloat[FAM]{\includegraphics[scale = 0.065]{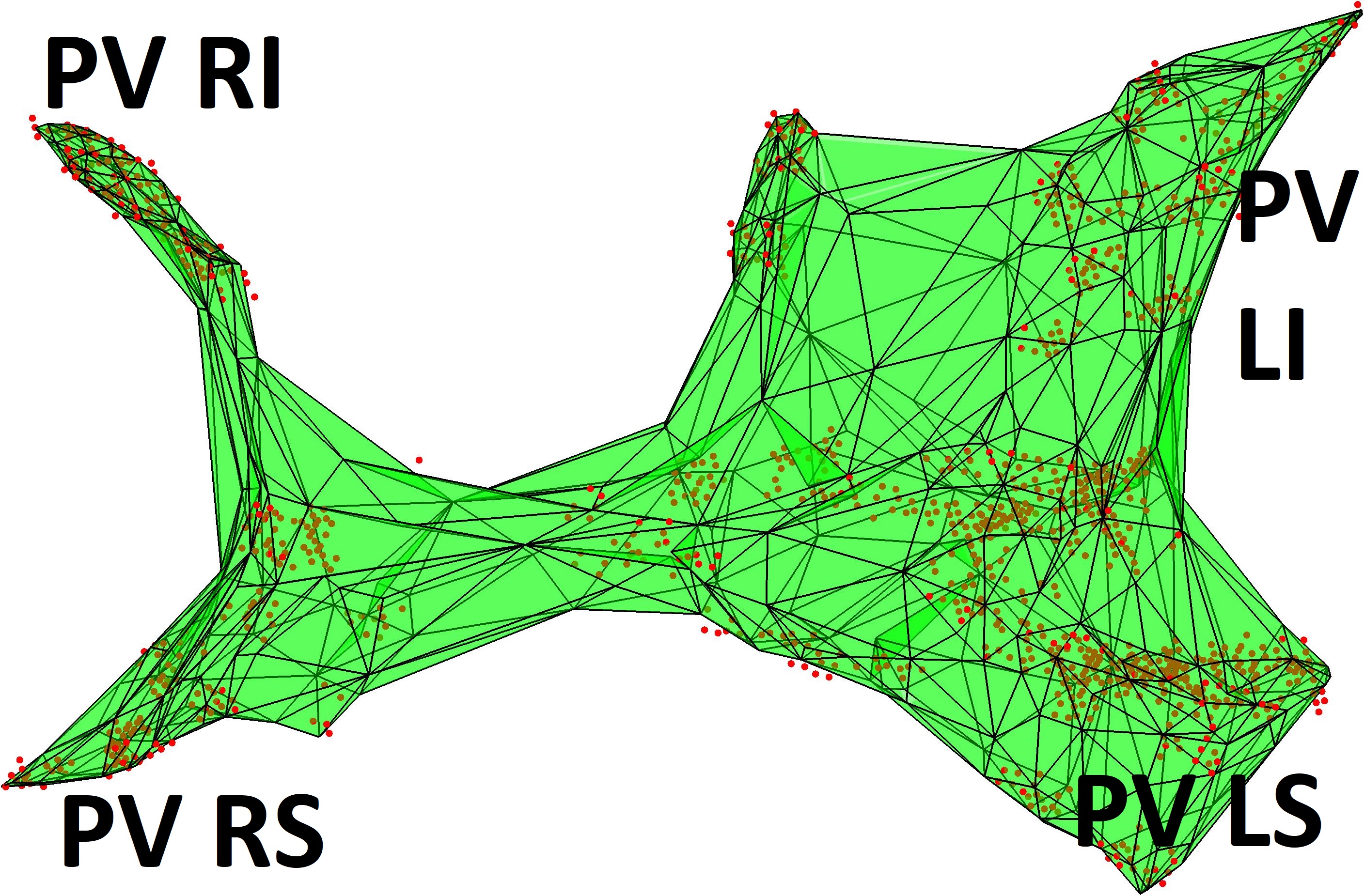}} \qquad
  \subfloat[\ded]{\includegraphics[scale = 0.065]{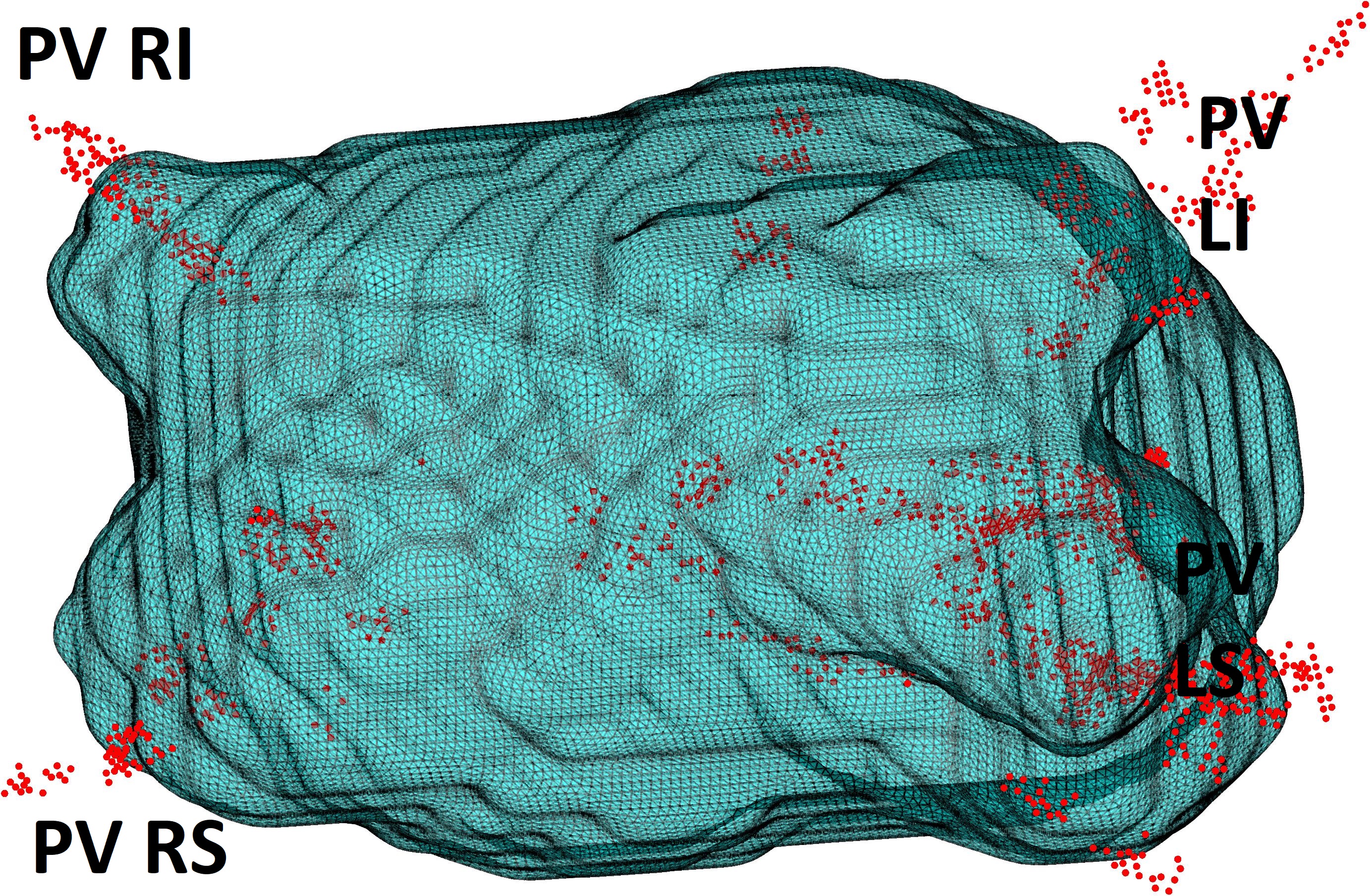}}
    \caption{Comparing FAM and \dedsp reconstructions using the path of the initial bearing. Note that an anatomically correct solution is obtained only for \ded.}
\label{fig:eect_fam_vs_rec}
\end{figure}

Qualitative results are shown in the next few Figures. A comparison of the \dedsp network reconstruction to the mean shape is shown in Figure \ref{fig:eect_swr_macmp_a} and Figure \ref{fig:eect_swr_macmp_all}. Visualizations of \vnetsp  reconstruction with anatomical issues are presented in Figure \ref{fig:eect_vnet_iregular}. Finally, we exemplify the advantage of using the proposed \dedsp solution over the existing FAM system - for the 3 minutes point-set acquisition interval in Figure \ref{fig:eect_fam_vs_rec}.

Figure \ref{fig:eect_swr_macmp_a} shows a
comparison of reconstruction results focusing on the location and orientation of the PVs, for 2 sample cases. \dedsp reconstruction is shown on top and  the mean shape reconstruction is shown on the bottom.  CT ground truth is shown in yellow, and the reconstruction  results are in blue. Inspection reveals that the \ded \space network improved the estimated \PV s location and orientation. In particular, the mean shape reconstruction misses parts of the PVLS and PVRI.   

Additional examples are shown in Figure \ref{fig:eect_swr_macmp_all}. Here, four reconstruction methods (2 \dedsp network solutions, \vnet, and the mean-shape reconstruction) are compared - showing the reconstruction as well as the surface-to-surface distances. Reconstruction results are shown in odd-numbered rows, with the ground-truth CT surfaces shown in even-numbered rows. On top of the surfaces, We present distance maps for a $15mm$ radius from PV ostia points and  unbounded  (as in Table \ref{exprPathRecCTTable}). Distance maps measured from reconstruction to CT, are presented in odd-numbered rows, and a reverse visualization, from CT to reconstruction is shown in the even-numbered rows.   
We  note that the errors are larger  in the mean-shape reconstruction - with fewer dark blue regions (overall lighter color-scale).  The right \PV s are magnified to support the visualization of these regions in the three networks compared to the mean shape on the bottom. 

In many of the cases experimented with (including many samples shown), the use of the   \vnet \space solution resulted in a smooth  reconstructed surface and gave low quantitative errors. However, a closer look at clinical examples reveals the presence of anatomical deformation in the reconstruction. 
Figure \ref{fig:eect_vnet_iregular} displays a few such cases 
Three examples are shown in which anatomical structures are added or removed from the LA surface, as confirmed by the given CT anatomy. Similar, minor to severe anomalies, were found in more than half of the 26 CT cases, thus making the value of \vnet \space for this task questionable.

Figure \ref{fig:eect_fam_vs_rec} compares the reconstruction results for FAM vs \dedsp solution after acquiring the  initial (3-minute interval) bearing path. In this case the FAM reconstruction is not anatomically viable while the \dedsp reconstruction is anatomically correct.


 \section{Discussions}
 \label{sec:discussions}
Our findings suggest that a neural-network based solution is capable of reconstructing the LA shape from a very sparse point cloud path.
Although the well-known \vnet \space network  exhibited superior performance in many of the quantitative metrics, the reconstructions contained many poorly represented parts of the anatomy, thus making it unsuitable as a clinical solution.
Thus, the \ded \space network regularized by a novel spatial weight smoothing and boundary enhancement map was the overall best performing architecture. The \dedsp network solution learned smooth probable atrium parts and combined them to recover a  realistic-looking left atrium.


Using real clinical cases enabled a clearer distinction of network performance: When quantifying the performance error, we saw that the DED solution was consistent across various values for the SWR parameter; The reconstruction accuracy was not far behind the FAM performance and achieved statistically significant mean error improvement over the mean shape for the clinically relevant regions (CT evaluation), and contact points. The focus on clinically relevant regions and contact points is critical, since global metrics for the entire shape such as DICE, AVDist, and surface-to-surface distance, accumulate errors from many volume/surface points over the LA body. These constitute large regions with a small number of measurements that are subject to large variations due to registration and motion errors. Methods that fit well over the data could improve these global metrics while still generating ill-formed anatomy.     


Note the inherent trade-off between fitting the input data and preserving its resemblance to the learned atrium anatomy. 
For the current task, we prefer networks that do not over-fit the data. This is likely to hold even for inputs that exhibit less resemblance to the data in the training set as seen in the clinical cases. \ded \space without SWR introduced large uncertainty regions and non-anatomical structures to cope with paths less similar to those trained upon. 

The DED solution proposed combines two time intervals: time to acquire the path, and time for inference in which the LA is reconstructed.
The first, path acquisition interval, is approximately 3 minutes. This  enables an anatomically correct visualization with reasonable accuracy within a third of the acquisition time, as compared to current methods in the clinic (FAM). This early visualization simplifies the procedure, supports novel workflows, eases operator load, and may contribute to safety, especially for inexperienced physicians. The  \ded \space network inference is conducted in  real-time. 

Qualitative inspection clearly revealed that the \vnet \space solution was not anatomically probable. It produced a smooth but twisted shape. 
This might be due to the lack of global understanding of the shape in this network since each convolution operation is local.
This outcome is consistent with the findings in \cite{Xiong2022} in which the authors show that their \vnet-based solution  performance depends  on a large percent coverage of the atria surface  (note that this requires a longer acquisition time, as in FAM). 



 \section{Conclusion}
 \label{sec:conclusions}
In this study, we used a neural network-based approach to recover the shape of the left atrium from catheter paths. We used an anatomically based model to generate left atrium shapes and a novel algorithm to create paths inside them. These paths were sufficient to teach the network to complete the left atrium shape from actual catheter paths in human cases.


The current work is part of an effort to build a system that infers the most probable anatomically correct reconstruction while considering currently available data. 
The data acquired in the clinic may contain additional information such as force measurements, electrocardiograms that hint at the anatomical region, etc. Models that can incorporate additional data and user inputs regarding the anatomy will further improve and simplify the mapping process.

The network was trained for the widespread anatomy of four \PV s, and typical size variants. To deal with variations in anatomy, a model trained with the appropriate dataset should be used. 
The network approach, as proposed in this work, is a general approach that is likely to be applicable to other heart chambers and  organs, different imaging techniques, and different types of partial inputs.  

\appendix
\setcounter{section}{0}
\renewcommand{\thesection}{\Alph{section}}
\section{Appendix}

\label{sec:app}
This section presents the details of the synthetic path creation algorithm.
\subsection{Marking points in the PV's interior for all atria}
\label{sec:synpathstg1}
In this stage, for each input atrium, we found a point for every \PV \space that is inside the ostium. We utilized the mean LA shape reference, having manually selected a point on each artificially closed \PV. The corresponding points on all the other LA samples were then generated by taking advantage of the fact that all atrial surfaces are similarly produced by the LA instance generator. These samples have Gaussian-like distribution for the orientations and the locations of the \PV s. This step operated over the triangulated mesh representation to utilize the geometric properties of the surface. In the mesh representation, the chosen landmark points of the mean shape are vertices of the mesh.
\begin{figure}
 	\centering
 	\subfloat[]{\includegraphics[scale=0.30]{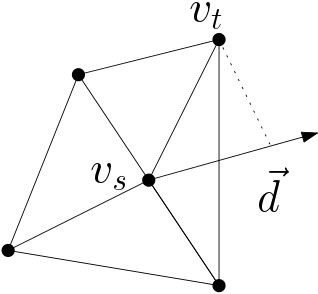}} \qquad	\subfloat[]
{\includegraphics[scale=0.2]{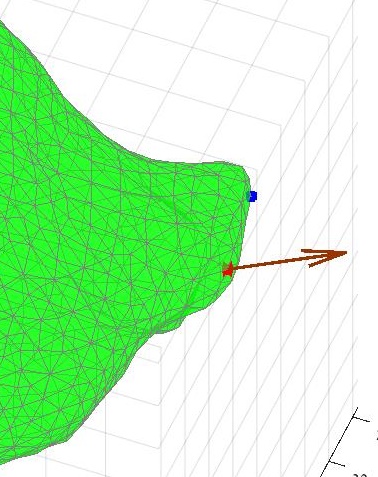}}
 	\caption{Illustration of local PV search. We choose vertices in the direction that matches PV orientation until no improvement can be made. (a) The direction $\vec{d}$ points from $v_s$ to $v_t$ as the next step. (b) The red Arrow shows the general orientation of the PV that should be followed. }
    \label{fig:pathCreation}
 \end{figure}
 
 Given the input atrium, we found for each landmark vertex (of the mean shape),  the closest vertex over the input LA mesh.  
 For each vertex, the normal (to the surface) at the abovementioned vertex of the mean atria is used as a direction to 'slide' across the new atrium vertices from the initial vertex until we can no longer advance in the direction of the normal, as seen in Figure \ref{fig:pathCreation}. This vertex is likely to be on the same \PV \space in the input atrium.  

As shown in Algorithm \ref{alg:the_alg} the input to the algorithm consists of the triangulated mesh (vertices and triplets of face membership) of the new atrium, a vertex $p$ which is the original vertex in the mean atrium, a direction $\vec{d}$ which is the normal at $p$ in the mean atrium and a threshold $\epsilon$. In line \ref{line:curr_vertex} we assign the nearest vertex on the new atrium to the initial vertex from the mean atrium. Lines \ref{line:vstart} - \ref{line:vend} compute the direction vectors from the current vertex to its neighbors. In Lines \ref{line:diffstart} - \ref{line:diffend} we assign the neighboring vertex whose direction vector projection over $\vec{d}$ is maximal. The loop terminates once the projected step length is less than $\epsilon$.

\SetKwRepeat{Do}{do}{while}%
\begin{algorithm}

\SetAlgoLined
\SetKwFunction{findPointInPv}\textsc{FindPointInPV}
\Indm\findPointInPv{$mesh,p,\overrightarrow{d},\epsilon$}\\ 
\Indp $current\_vertex \gets FindNearestVertex(mesh,$p$)$ \label{line:curr_vertex}\\ 
\Do{$direction\_ difference \geq \epsilon$}
{$neighbouring\_ vertices \gets getNeighbours(mesh,current\_vertex)$ \label{line:vstart}\\
 $\overrightarrow{dist\_vectors}  \gets neighbouring\_ vertices-current\_vertex$\\
 $\overrightarrow{dist\_vectors} \gets \frac{dist\_vectors}{\| dist\_vectors\|}$ \label{line:vend} \\
 $direction\_ difference \gets \max{\overrightarrow{dist\_vectors}\cdot \vec{d} } \label{line:diffstart}$\\
 $current\_vertex \gets arg\max_{v \in neighbouring\_ vertices}{\overrightarrow{dist\_vectors}\cdot \vec{d}} $} \label{line:diffend}
 \caption{Find \PV{} Algorithm}
 \label{alg:the_alg}
\end{algorithm}

The septum point is a bit different to locate as it is chosen clinically by the physician on the septal wall between the right and the left atria. We decided to simulate this by finding the nearest vertex to a chosen septum in the mean shape and then sampled a vertex (using a Gaussian distribution) around this vertex.

\subsection{Creating a Path between two PV's }
\label{sec:ptppath}
\begin{figure}[]
 	\centering
 	\subfloat[]{\includegraphics[scale=0.12]{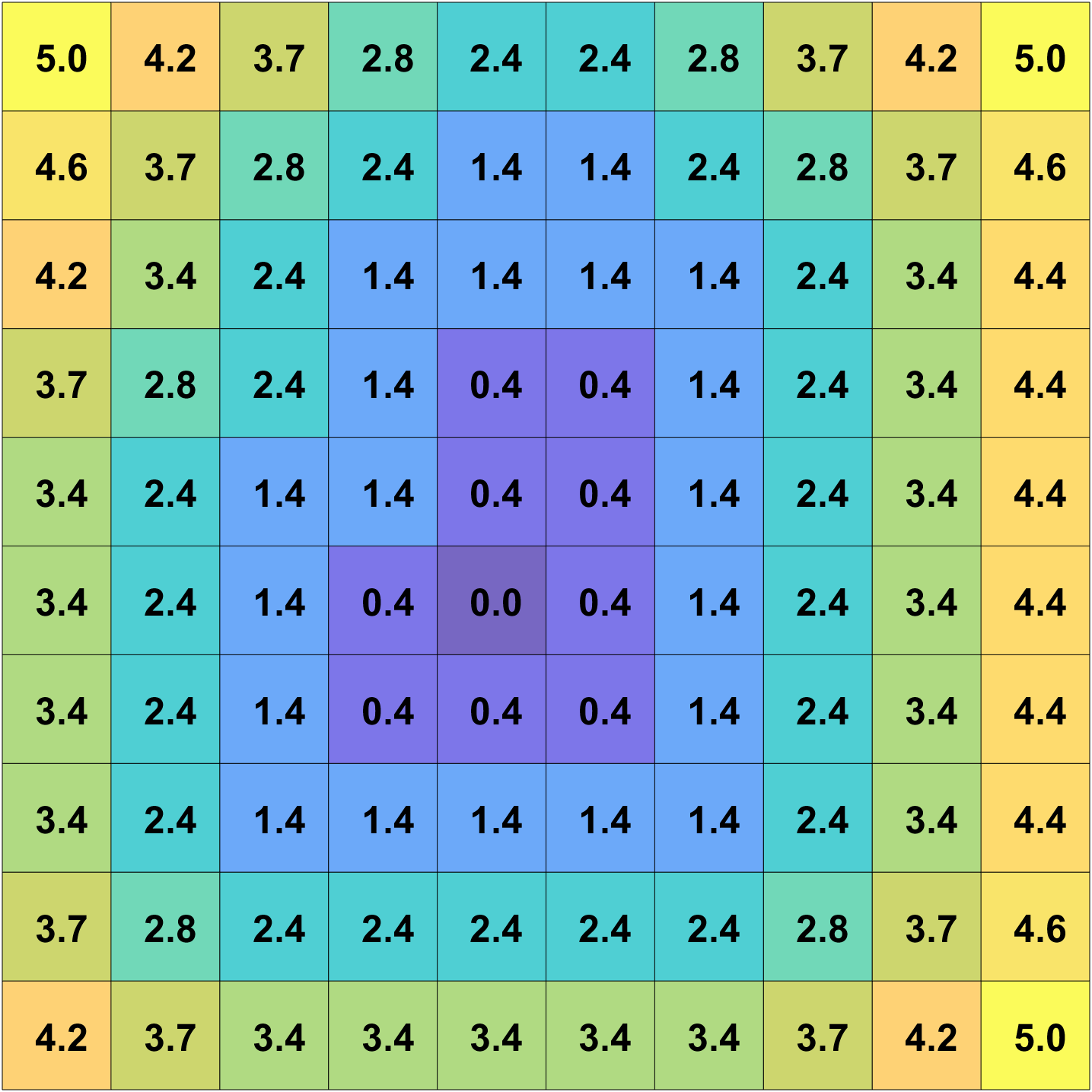}\label{fig:vertexcreateweight}}   \subfloat[]
{\includegraphics[scale=0.50]{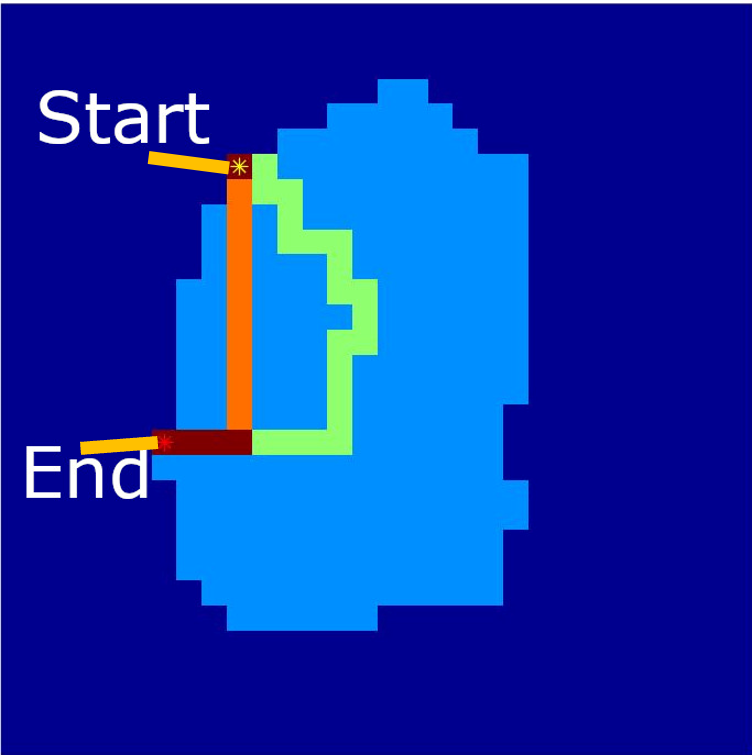}\label{fig:pathpower}}
 	\caption{Path Weighting. (a) Cost of a vertex on the graph. The weight of the edge is the mean of its vertices. (b) $\alpha$ effect over the path. The orange path has $\alpha=0$, and  $\alpha=1$ for the green path. The red area is common to both paths. }
    \label{fig:pathweightings}
 \end{figure}

A catheter maneuver between two points inside the left atrium is very different from a straight line due to the limited degrees of freedom of the catheter. 
A realistic path between two \PV s \space  is relatively short and lies close to the center of the atrium (since a retraction to the center is usually performed while going from one area to the other). We express this problem by using a graph weighted such that the optimal path between two nodes will follow these two considerations. 
First, we use the discrete volume representation where voxels inside the atrium are represented by one. Next, we find the voxels that represent the two \PV s. The graph is built such that each voxel in the volume is a node edge connected to its six neighbors. The boundary of the volume is extracted and a discrete signed distance transform (Euclidean distance) is computed over the volume. To compute this metric, a set of boundary voxels are selected having a distance of zero, and all other voxels are assigned a value that is the distance to the nearest boundary voxel. The value is negative for voxels outside the atrium, and positive inside. Thus, voxels close to the artium center have high Euclidean distance values.
To define a cost for minimization, we denote the maximal distance as $m_w$ and assign each voxel the weight $m_w-w_{dt}$ where $w_{dt}$ is the voxel distance transform, as seen in Figure \ref{fig:vertexcreateweight}. The edge between two neighboring voxels is assigned the mean of their weights. Edge cost increases as we move further away from the atrium's center. The shortest path between any two \PV s is found using the Dijkstra algorithm \cite{cormen}, with regard to our objectives favoring a short path (since each edge adds weight) that is close to the atrial center. We also define an $\alpha$ parameter where each weight $w$ becomes $w^\alpha$. When $\alpha$ reaches zero the path tends toward the shortest path (as each weight becomes one), whereas $\alpha$ progresses towards one and beyond the path tends towards the center, see Figure \ref{fig:pathpower}.

\subsection{Path Part Integration}
\label{sec:pathintegration}
In order to create a full synthetic path we first we project the septum point from the template sample to the current sample.
We then find point to point paths as described in Appendix \ref{sec:ptppath}. The first path is from the septum to the left superior, the second continues to left inferior, the third to the right inferior and the last one to right superior; see Figure \ref{fig:pathpartscol}.
Different parts of the path can be equilibrated by using a different power squashing ($\alpha$) for each part, depending on how much we have to pull the catheter back to complete the path.
The $\alpha$ parameter for each path was determined experimentally. The chosen values were $[0.001,4,1,4]$ in the presented order. A sample path is shown in Figure \ref{fig:examplepath}.


\bibliographystyle{IEEEtran}
\bibliography{sample.bib}

%
%






\end{document}